\documentclass[reprint,preprintnumbers,
amsmath,amssymb,aps,nofootinbib,superscriptaddress]{revtex4-1}
\usepackage[pdftex]{graphicx}
\usepackage{dcolumn}
\usepackage{bm}
\usepackage{cases}
\usepackage{natbib}
\usepackage[pdftex]{hyperref}
\usepackage{color}
\definecolor{phthaloblue}{rgb}{0.0, 0.06, 0.54}
\definecolor{bluscuro}{rgb}{0.15, 0.2, .85}
\definecolor{rossos}{cmyk}{0,1,1,0.55}
\definecolor{bluchiaro}{cmyk}{1,.3,0.,0.1}
\hypersetup{
    colorlinks=true,
    linkcolor=bluscuro,
    citecolor=phthaloblue,
    filecolor=phthaloblue,
    urlcolor=phthaloblue,
    }
\makeatletter \def\@eqnnum{{\normalsize \normalcolor (\theequation)}}  
\makeatother %
\newcommand{\para}[1]{\par\vspace{2mm}\noindent\textbf{\emph{{#1}}.---}}

\begin{document}
\preprint{\begin{minipage}[b]{1\linewidth}
\begin{flushright}Preprint numbers: KEK-TH-1989, KEK-Cosmo-208\end{flushright}
\end{minipage}}
\title{Electroweak Vacuum Collapse induced by
Vacuum Fluctuations of the Higgs Field
 around Evaporating Black Holes}
\author{Kazunori Kohri}
\email{kohri@post.kek.jp}
\affiliation{KEK Theory Center, IPNS, KEK, Tsukuba, Ibaraki 305-0801, Japan}
\affiliation{The Graduate University of Advanced Studies (Sokendai),Tsukuba, Ibaraki 305-0801, Japan}
\affiliation{Rudolf Peierls Centre for Theoretical Physics,
The University of Oxford, 1 Keble Road, Oxford, OX1 3NP, UK}
\author{Hiroki Matsui}
\email{matshiro@post.kek.jp}
\affiliation{KEK Theory Center, IPNS, KEK, Tsukuba, Ibaraki 305-0801, Japan}
\affiliation{The Graduate University of Advanced Studies (Sokendai),Tsukuba, Ibaraki 305-0801, Japan}

\begin{abstract}
In this paper, we discuss the Higgs vacuum stability around 
evaporating black holes.
We provide a new approach to investigate 
the false vacuum decay around the black hole and 
clearly show how vacuum fluctuations of the Higgs
induce a gravitational collapse of the vacuum.
Furthermore, we point out that the backreaction of the Hawking
radiation can not be ignored and the gravitational vacuum decay 
is exponentially suppressed. However, a large number of the evaporating
(or evaporated) primordial black holes threaten the existence of the Universe
and we obtain a new upper bound on the evaporating PBH abundance 
from the vacuum stability.
Finally, we show that the high-order corrections of the BSM or QG would not 
destabilize the Higgs potential, otherwise 
even a single evaporating black hole triggers
a collapse of the electroweak vacuum. 
\end{abstract}
\date{\today}
\maketitle
\flushbottom
\allowdisplaybreaks[1]

%%%%%%%%%%%%%%%%%%%%%%%%%%%%%%%%%
%%%%%%%%%%%%%%%%%%%%%%%%%%%%%%%%%
\section{Introduction}
\label{sec:intro}
%%%%%%%%%%%%%%%%%%%%%%%%%%%%%%%%%
In the late 1970s, Hawking~\cite{Hawking:1974sw} showed that black
holes emit thermal radiation at the Hawking temperature
$T_{\rm H}=1/(8\pi M_{\rm BH})$ due to quantum particle creation 
on strong gravitational field
%%%%%%%%%%%%%%%%%%%%%%%%%%%%%%%%%
\footnote{In the present paper we use natural units where $\hbar =c=k_{B}= G =1$.}
%%%%%%%%%%%%%%%%%%%%%%%%%%%%%%%%%
where $M_{\rm BH}$ is the black-hole mass.  
The quantum particle creation around the black hole determines the fate of the
evaporating black hole which is still unknown and closely related with
the information loss puzzle~\cite{Hawking:1976ra}, and furthermore,
leads to the spontaneous symmetry restoration~\cite{Hawking:1980ng} or
the false vacuum decay around the black
hole~\cite{Hiscock:1987hn,Berezin:1987ea,Arnold:1989cq} which bring
cosmological singular possibility. 
The latter case can also be interpreted as the catalysis 
induced by vacuum fluctuations around the black hole.

One of the most curious puzzles of the observed Higgs boson
at the LHC experiment~\cite{Aad:2015zhl,Aad:2013wqa,
Chatrchyan:2013mxa,Giardino:2013bma} is that 
the effective Higgs potential develops an instability at the 
scale $\Lambda_{I} \approx 10^{11}\ {\rm GeV}$
%%%%%%%%%%%%%%%%%%%%%%%%%%%%%%
%%%%%%%%%%%%%%%%%%%%%%%%%%%%%%%%%%%%%%
\footnote{
The instability scale $\Lambda_{I}$ is generally determined by the value of the Higgs boson mass $m_{h}$ and the top quark mass $m_{t}$.
The current observed masses of the Higgs boson $m_{h}=125.09 \pm 0.21 ({\rm stat}) \pm 0.11({\rm syst})\ {\rm GeV}$~\cite{Aad:2015zhl,Aad:2013wqa,
Chatrchyan:2013mxa,Giardino:2013bma} and
the top quark $m_{t}=172.44\pm 0.13 ({\rm stat}) \pm 0.47 ({\rm syst})\ {\rm GeV}$~\cite{Khachatryan:2015hba} suggest the instability scale to be 
$\Lambda_{I} \approx 10^{11}\ {\rm GeV}$~\cite{Buttazzo:2013uya} although this instability scale $\Lambda_{I}$ has the gauge dependence
(see~\cite{DiLuzio:2014bua,Andreassen:2014eha,Andreassen:2014gha,Lalak:2016zlv,Espinosa:2016uaw,Espinosa:2016nld} for the detail).}
%%%%%%%%%%%%%%%%%%%%%%%%%%%%%%%
%%%%%%%%%%%%%%%%%%%%%%%%%%%%%%%%%%%%%
where we assume no corrections of the beyond Standard Model (BSM) and the quantum gravity (QG)~\cite{Branchina:2013jra,Lalak:2014qua,Branchina:2014usa,Branchina:2014rva}.
Thus, the current electroweak vacuum state of the Universe is not stable, 
and finally collapses through quantum 
tunneling~\cite{Kobzarev:1974cp,Coleman:1977py,Callan:1977pt}
although the timescale of the decay is longer than the age of our Universe~\cite{Degrassi:2012ry,Isidori:2001bm,Ellis:2009tp,EliasMiro:2011aa}.
However, vacuum fluctuations induced by the strong gravitational field 
drastically change the stability of the electroweak vacuum.
%%%%%%%%%%%%%%%%%%%%%%%%%%%%%
%%%%%%%%%%%%%%%%%%%%%%%%%%%%%
\begin{figure}[t]
\includegraphics[width=90
mm]{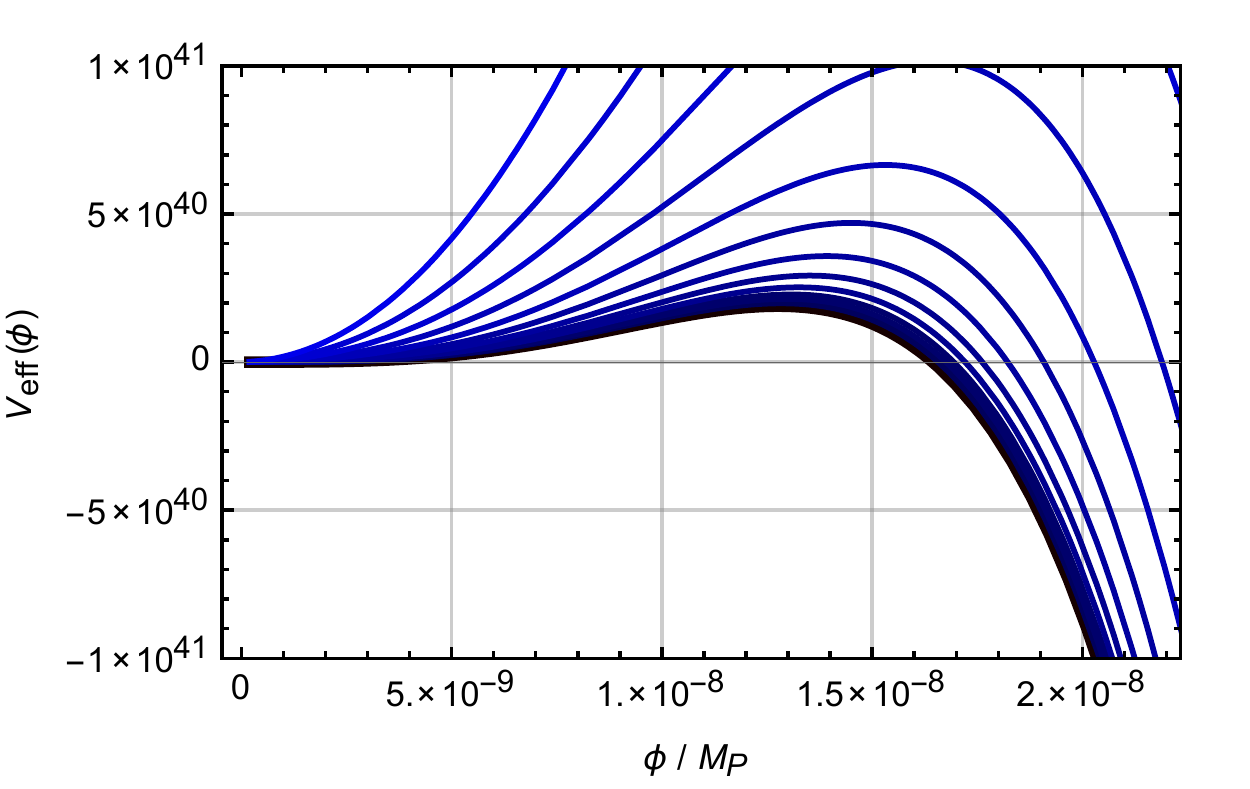}
\caption{The schematic diagram of the effective Higgs potential 
around evaporating black hole. The back-reaction of the Hawking radiation 
around the black hole stabilizes the Higgs potential 
at $T_{\rm H}\gtrsim \Lambda_{I}$.
The figure sets the present best-fit values of $m_{h}$ and $m_{t}$,
and takes different Hawking temperatures $T_{\rm H}=10^{9.0}-10^{10.2}\ {\rm GeV}$
for the Higgs potential on Eq.~(\ref{eq:thamlhiggs}). }
\label{Fig:blackhole2}
\end{figure}
%%%%%%%%%%%%%%%%%%%%%%%%%%%%%%
%%%%%%%%%%%%%%%%%%%%%%%%%%%%%%
The recent works suggest that the electroweak vacuum is generally unstable 
on various gravitational or cosmological backgrounds, e.g. 
during inflation~\cite{Espinosa:2007qp,Herranen:2014cua,Hook:2014uia,Kearney:2015vba,Kohri:2016qqv,Kohri:2017iyl,Czerwinska:2016fky,Gong:2017mwt,East:2016anr,Espinosa:2015qea,Joti:2017fwe}  corresponding to de-Sitter spacetime, 
after inflation~\cite{Herranen:2015ima,
Kohri:2016wof,Ema:2016kpf,Enqvist:2016mqj,Postma:2017hbk,Ema:2017loe} and 
around evaporating black holes~\cite{Burda:2015isa,Burda:2015yfa,
Burda:2016mou,Grinstein:2015jda,
Tetradis:2016vqb,Cheung:2013sxa,Canko:2017ebb}.
In particular, recent discussion about the electroweak
vacuum stability around evaporating black holes has been growing.  
The black holes emitting the Hawking radiation reduce the mass and 
finally evaporate. At the final stage of evaporating black holes,
the Hawking temperature is extremely high and 
the influence of the vacuum stability can not be ignored~\cite{Cheung:2013sxa,Canko:2017ebb,Gorbunov:2017fhq,Mukaida:2017bgd}.
Generally, there are no mechanisms to prevent the formation
of such small black holes which finally evaporates during the history of the Universe.
In particular, the primordial black holes (PBH) formed 
by large density fluctuations in the early
Universe~\cite{Hawking:1971ei,Carr:1974nx,Carr:1975qj} are
incompatible with the Higgs vacuum stability~\cite{Gorbunov:2017fhq,Cole:2017gle}.  
These phenomena could imposed serious constraints on the cosmology 
or the Particle Physics beyond the Standard Model.

However, there is some controversy about how 
evaporating black holes trigger off the Higgs vacuum decay.
The analysis~\cite{Burda:2015isa,Burda:2015yfa,Burda:2016mou}
of the false vacuum decay around black holes 
by the Coleman-De Luccia (CDL) instanton
formalism~\cite{Coleman:1980aw} obscure the dependence of 
the quantum particle creation around the black hole.
Furthermore, the back-reaction of the
Hawking radiation with $T_{\rm H}=1/ (8\pi M_{\rm BH})$ can not be ignored
as the de Sitter spacetime~\cite{Hawking:1981fz},
and it is reasonable intuitively to assume the
Higgs potential~\cite{Arnold:1991cv,Carrington:1991hz,
 Anderson:1991zb,Delaunay:2007wb,Rose:2015lna} at finite temperature instead of the
zero-temperature in the environment of the thermal Hawking flux.
The thermal corrections can generally stabilize the
Higgs potential (see Fig.~\ref{Fig:blackhole2}), and the possibility of the Higgs 
vacuum decay around black holes would be expected to be 
lower than what was considered.
Besides the backreaction issues of the Hawking radiation, 
the CDL instanton formalism is physically obscure, 
and therefore, one needs an another derivation of the decay rate 
without relying on instanton methods
%%%%%%%%%%%%%%%%%%%%%%%%%%%%%
%%%%%%%%%%%%%%%%%%%%%%%%%%%%%
\footnote{The stochastic approach of the false vacuum decay 
in the black hole background was investigated by~\cite{Iso:2011gb}.}.
%%%%%%%%%%%%%%%%%%%%%%%%%%%%%
%%%%%%%%%%%%%%%%%%%%%%%%%%%%%
In de Sitter space it is well known that one can get the probability of tunneling 
to true vacuum by two-point
correlation function $\left< { \delta \phi }^{ 2 } \right>$~\cite{Linde:1993xx}.
The decay probability using the two-point correlation
function which describes the 
vacuum fluctuations in quantum field theory (QFT) 
precisely match the Euclidean correct results~\cite{Linde:1991sk,
Linde:1978px,Dine:1992wr}.

In accordance to this sprit we investigate how vacuum fluctuations 
around black holes trigger off the Higgs vacuum decay by using the 
two-point correlation function $\left< { \delta \phi }^{ 2 } \right>$.
We discuss the renormalized expression of the correlation function 
$\left< { \delta \phi }^{ 2 } \right>_{\rm ren}$~\cite{Candelas:1980zt} 
for three possible vacua which are Boulware, Unruh and Hartle-Hawking vacuum
in Schwarzschild spacetime, and confirm that 
the Unruh vacuum is an appropriate vacuum which 
describes evaporating black holes.
By using the renormalized two-point correlation function 
$\left< { \delta \phi }^{ 2 } \right>_{\rm ren}$ in Unruh vacuum 
we provide a new approach to investigate the false vacuum decay 
around evaporating black holes and apply the Higgs vacuum stability.
Furthermore we clearly show that the backreaction of the Hawking radiation
can not be physically ignored and the vacuum fluctuations stabilize the 
effective Higgs potential. Finally, based on the consideration, we discuss
cosmological constraints on the PBHs due to the vacuum stability of the Higgs.
We find that just one evaporating PBH does not cause 
a collapse of the electroweak vacuum, but on the other hand 
a large number of PBHs is real serious.
In this paper, we provide a quantitative description of 
the Higgs vacuum stability around evaporating black holes and 
get a new bound of the PBHs.

The organization of this paper is as follows. In Sec. \ref{sec:vacuum}, we introduce
renormalized vacuum fluctuations for various vacua in Schwarzschild spacetime. 
In Sec. \ref{sec:block holes}, we argue how vacuum fluctuations around 
evaporating black holes modify the Higgs potential.
In Sec. \ref{sec:electroweak}, we discuss stabilities of the electroweak vacuum around
the block hole and give a cosmological constraint of the PBHs.
Sec. \ref{sec:conclusion} is devoted to our conclusions 
and future outlooks.

%%%%%%%%%%%%%%%%%%%%%%%%%%%%%%%%%%%
%%%%%%%%%%%%%%%%%%%%%%%%%%%%%%%%%%%
\section{The vacuum fluctuations around Schwarzschild black hole}
\label{sec:vacuum}
%%%%%%%%%%%%%%%%%%%%%%%%%%%%%%%%%%%
%%%%%%%%%%%%%%%%%%%%%%%%%%%%%%%%%%%
Formally, quantum effects of the vacuum fluctuations are
described by the vacuum expectation value of the energy momentum
tensor $\left< { T }_{\mu \nu} \right>$ or the correlation
function $\left< { \delta \phi }^{ 2 } \right>$ in the quantum field
theory (QFT) in curved spacetime.  
The former $\left< { T }_{\mu \nu} \right>$ provides a exact description of the
quantum back-reaction on the geometry,
and it is crucial to determine the stability of the background spacetime 
and the fate of the evaporating black hole.
The two-point correlation function $\left< {  \delta \phi  }^{ 2 } \right>$ 
corresponds to the vacuum fluctuation and
plays essential role in the vacuum stability.
But the correlation function usually diverges and unphysical divergences must be 
eliminated by the regularization and the renormalization method.

The renormalization of 
vacuum fluctuations for massless scalar field in Schwarzschild spacetime
has been well-know and analytical estimation is possible.
However, the massive case requires complicated numerical calculations.
For briefness, in this section, we review the renormalized vacuum fluctuations 
for the massless case in Boulware, Unruh and Hartle-Hawking vacuum 
following~\cite{Candelas:1980zt} and discuss which is an appreciate vacuum state 
to describe the evaporating black hole.

%%%%%%%%%%%%%%%%%%%%%%%%%%%%%
%%%%%%%%%%%%%%%%%%%%%%%%%%%%%
\begin{figure}[t]
\includegraphics[width=85mm]{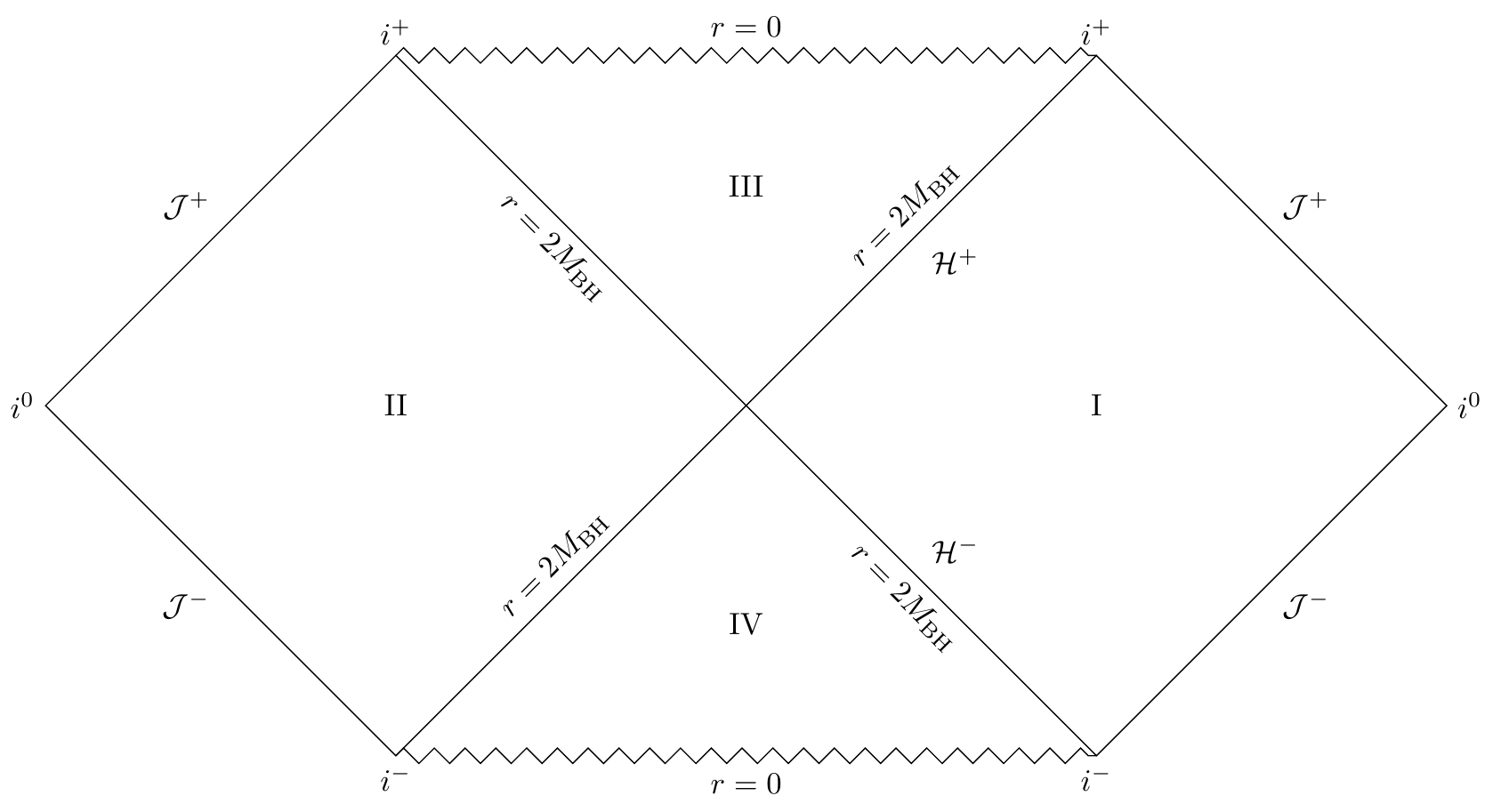}
\caption{The Penrose-Carter diagram of the maximally extended Schwarzschild manifold. 
Regions I or II are asymptotically flat, Region III is the black hole, and Region IV is the white hole.
$\cal{H}^{+}$ corresponds to the (future) black hole horizon, $\cal{H}^{-}$ 
is the (past) black hole horizon, $\cal{J}^{+}$ corresponds to the (future) null infinity
and $\cal{J}^{-}$ is the (past) null infinity.}
\label{Fig:PCD}
\end{figure}
%%%%%%%%%%%%%%%%%%%%%%%%%%%%%%
%%%%%%%%%%%%%%%%%%%%%%%%%%%%%%

The metric in the Schwarzschild coordinates (where we ignore 
the quantum backreaction of the scalar field on the geometry) can be written by
\begin{align}
d{ s }^{ 2 }=\ &-\left( 1-\frac { 2M_{\rm BH} }{ r }  \right) d{ t }^{ 2 }+\frac { d{ r }^{ 2 } }{ 1-2M_{\rm BH}/r }\nonumber 
\\ &+{ r }^{ 2 }\left( d{ \theta  }^{ 2 }+\sin ^{ 2 }{ \theta  } d{ \varphi  }^{ 2 } \right) \label{eq:dshdh},
\end{align}
which covers the exterior region $r>2M_{\rm BH}$ of the spacetime 
where $M_{\rm BH}$ is the black hoe mass.
The above singularity at the horizon $r = 2M_{\rm BH}$
can be removed by transforming to Kruskal coordinates.
By taking the Kruskal coordinates, we obtain the following metric
\begin{align}
d{ s }^{ 2 }=\ &\frac { 32{ M }^{ 3 }_{\rm BH} }{ r } { e }^{ -r/2M_{\rm BH} }dUdV \nonumber
\\ & +{ r }^{ 2 }\left( d{ \theta  }^{ 2 }+\sin ^{ 2 }{ \theta  } d{ \phi  }^{ 2 } \right) \label{eq:dshdhdh},
\end{align}
where these coordinates $U$ and $V$ are formally given by
\begin{align}
U&=-M_{\rm BH}{ e }^{ -u/4M_{\rm BH}},\quad V=M_{\rm BH}{ e }^{ v/4M_{\rm BH} },\nonumber \\
u&=t- r- 2M_{\rm BH}\ln { \left( \frac { r }{ 2M_{\rm BH} } -1 \right)  },  \\
v&=t + r + 2M_{\rm BH}\ln { \left( \frac { r }{ 2M_{\rm BH} } -1 \right)  }.\nonumber
\end{align}
The Schwarzschild coordinates of Eq.~(\ref{eq:dshdh})
cover only a part of the spacetime, whereas 
the Kruskal coordinates of Eq.~(\ref{eq:dshdhdh}) cover the extended spacetime and 
is regular at the black hole horizon.
These features of the Schwarzschild geometry 
are summarized in the Penrose-Carter diagrams as Fig.~\ref{Fig:PCD}.

In curved spacetime, there is no unique vacuum and 
we must take an appropriate vacuum state.
In the Schwarzschild spacetime, there are three well defined vacua, namely:
Boulware vacuum 
(vacuum state around a static star)~\cite{Boulware:1974dm,Boulware:1975pe},
Unruh vacuum (black hole evaporation)~\cite{Unruh:1976db}
and Hartle-Hawking vacuum (black hole in thermal equilibrium)~\cite{Hartle:1976tp} which 
correspond to the definitions of the normal ordering on the respective coordinates.

The Klein-Gordon equation for the massless scalar field $\phi \left( t,x \right) $ can be given by
\begin{align}
\left[ -{ \partial  }_{ \mu  }{ g }^{ \mu \nu  }\sqrt { -g } { \partial  }_{ \nu  }\right] \phi\left( t,x \right) =0,
\end{align}
where we drop the curvature term $\xi R \phi^{2}$  because 
the Ricci scalar becomes $R=0$ in Schwarzschild spacetime
%%%%%%%%%%%%%%%%%%%%%%%%%%%%%
%%%%%%%%%%%%%%%%%%%%%%%%%%%%%
\footnote{The Kretschmann scalar $K$ constructed of 
two Riemann tensors is non-zero, i.e $K=R_{abcd}R^{abcd}=48M^{2}/r^{6}$.}
%%%%%%%%%%%%%%%%%%%%%%%%%%%%%
%%%%%%%%%%%%%%%%%%%%%%%%%%%%%
for simplicity, but this approximation may brake down
when the quantum backreaction on the metric can not be neglected.
In the exterior region of Schwarzschild spacetime,
the scalar field $\phi \left( t,r,\theta,\varphi \right) $ is decomposed into the from
\begin{align}
&\phi\left( t,r,\theta,\varphi \right)  =\int _{ 0 }^{ \infty  }{d\omega  } \sum _{ l=0 }^{ \infty  }\sum _{ m=-l }^{ l }
\biggl( { a }_{ \omega lm }{ u }_{ \omega lm }^{in}  \\ & 
+{ a }_{ \omega lm }^{ \dagger  }{ u }_{ \omega lm }^{ in * }
 +{ b }_{ \omega lm } { u }_{ \omega lm }^{out}+{ b }_{ \omega lm }^{ \dagger  }
{ u }_{ \omega lm }^{ out * }  \biggr) \nonumber,
\end{align}
where these mode functions ${ u }_{ \omega lm }^{in}$ and 
${ u }_{ \omega lm }^{out}$
defines the vacuum state that ${ a }_{ \omega lm }\left|{ 0} \right>=
{ b }_{ \omega lm }\left|{ 0} \right>=0$ which corresponds to the boundary conditions.
In the Schwarzschild spacetime, these mode functions ${ u }_{ \omega lm }^{in}$ and 
${ u }_{ \omega lm }^{out}$ for the massless scalar field are given by
\begin{align*}
{ u }_{ \omega lm }^{in} & ={ \left( 4\pi \omega  \right)  }^{ -1/2 }{ R }_{ l }^{in} \left( r;\omega  \right) { Y }_{ lm }\left( \theta ,\varphi  \right) { e }^{ -i\omega t },\\
{ u }_{ \omega lm }^{out} & ={ \left( 4\pi \omega  \right)  }^{ -1/2 }{ R }_{ l }^{out} \left( r;\omega  \right) { Y }_{ lm }\left( \theta ,\varphi  \right) { e }^{ -i\omega t },
\end{align*}
where these radial functions ${ R }_{ l }^{in} \left( r;\omega  \right)$ and 
${ R }_{ l }^{out}  \left( r;\omega  \right)$ have the well-known asymptotic forms,
\begin{align*}
{ R }_{ l }^{in} \left( r;\omega  \right) \simeq \begin{cases}
     \begin{matrix} &{ B }_{ l }\left( \omega  \right) { r }^{ -1 }{ e }^{ -i\omega { r }_{ * } }\  \left(r\rightarrow 2M_{\rm BH} \right)
\\ &{ r }^{ -1 }{ e }^{- i\omega { r }_{ * } }+{ A }_{ l }^{in} \left( \omega  \right) { r }^{ -1 }{ e }^{ i\omega { r }_{ * } }\ \left(r\rightarrow \infty \right) \end{matrix} 
  \end{cases},
\end{align*}
\begin{align*}
{ R }_{ l }^{out} \left( r;\omega  \right) \simeq \begin{cases}
     \begin{matrix} &{ r }^{ -1 }{ e }^{ i\omega { r }_{ * } }+{ A }_{ l }^{out} \left( \omega  \right) { r }^{ -1 }{ e }^{ -i\omega { r }_{ * } }\  \left(r\rightarrow 2M_{\rm BH} \right)
      \\ &{ B }_{ l }\left( \omega  \right) { r }^{ -1 }{ e }^{ i\omega { r }_{ * } }\ \left(r\rightarrow \infty \right) \end{matrix} 
  \end{cases}.
\end{align*}
${ A }_{ l }^{in}  \left( \omega  \right)$, ${ A }_{ l }^{out}
\left( \omega  \right)$ and ${ B }_{ l }\left( \omega  \right)$
are the reflection and transmission coefficients~\cite{DeWitt:1975ys}.
The Boulware vacuum $\left|{ 0_{\rm B} } \right>$ is defined by taking ingoing and outgoing modes to be positive frequency 
with respect to the Killing vector $\partial_{t}$ of the Schwarzschild metric~\cite{Boulware:1974dm} and constructed by
using the scattering theory interpretation.

The two-point correlation function $\left< {  \delta \phi  }^{ 2 } \right>$ related with 
the Boulware vacuum $\left|{ 0_{\rm B} } \right>$ can be given by~\cite{Christensen:1977jc,Candelas:1980zt}:
\begin{align}
&\left< { 0_{\rm B} }|{ \delta { \phi  }^{ 2 }\left( x \right)  }|{ 0_{\rm B} } \right>= 
\frac{1}{16\pi^2}{ \int _{ 0 }^{ \infty  }{ \frac { d\omega  }{ \omega  }  }  } \nonumber \\ 
&\times \Biggl[ \sum _{ l=0 }^{  \infty }\left(2l+1\right)\left[
\left| { R }_{ l }^{in} \left( r;\omega  \right) \right|^{2}+
\left| { R }_{ l }^{out} \left( r;\omega  \right)\right|^{2}\Bigr] \right] \label{eq:asddg},
\end{align}
where the sum of these radial functions ${ R }_{ l }^{in} \left( r;\omega  \right)$ and 
${ R }_{ l }^{out} \left( r;\omega  \right)$ have the asymptotic forms,
\begin{align}
\sum _{ l=0 }^{  \infty }\left(2l+1\right)
\left|{ R }_{ l }^{in} \left( r;\omega  \right) \right|^{2} \sim \begin{cases}
\begin{matrix} &\frac{\sum_{ l=0 }^{ \infty  }{ \left( 2l+1 \right)  } 
{ \left| { { B }_{ l } }\left( \omega  \right)  \right|  }^{ 2 }}{4 M_{\rm BH}^{2}}\  \left(r\rightarrow 2M_{\rm BH} \right)
 \\ &4\omega^{2}\  \left( r\rightarrow \infty \right)   \end{matrix} 
\end{cases}, \nonumber
\end{align}
\begin{align}
\sum _{ l=0 }^{  \infty }\left(2l+1\right)
\left| { R }_{ l }^{out} \left( r;\omega  \right) \right|^{2} \sim \begin{cases}
\begin{matrix} &\frac{4\omega^{2}}{1-2M_{\rm BH}/r}\  \left(r\rightarrow 2M_{\rm BH} \right)  \\ & 
\frac{\sum_{ l=0 }^{ \infty  }{ \left( 2l+1 \right)  } { \left| { { B }_{ l } }\left( \omega  \right)  \right|  }^{ 2 }}{r^{2}}\  \left( r\rightarrow \infty \right)   \end{matrix} 
\end{cases}. \nonumber
\end{align} 
Thus, the two-point correlation function $\left< {  \delta \phi  }^{ 2 } \right>$ of Eq.~(\ref{eq:asddg})
has clearly divergences and must be regularized.
There are several regularization methods to eliminate the divergences 
in the quantum field theory (QFT), but the point-splitting regularization
is an extremely powerful and standard method to 
obtain the renormalized expression in curved spacetime. 
Let us consider temporarily $ { \delta \phi  }^{ 2 } \left(x\right)\rightarrow  { \delta \phi  }\left(x\right)
{ \delta \phi  }\left(x'\right)$ to remove the divergences and 
afterwards take the coincident limit $x' \rightarrow x$,
\begin{align}
\left< {  \delta \phi  }^{ 2 }\left(x\right) \right>_{\rm ren}=\lim _{ x'\rightarrow x }{ \left[\left< {  \delta \phi \left(x\right) }
{  \delta \phi \left(x'\right) }\right>-\left< {  \delta \phi \left(x\right) }{  \delta \phi \left(x'\right) }\right>_{\rm div} \right]  }, 
\end{align}
where $\left< {  \delta \phi \left(x\right) }{  \delta \phi \left(x'\right) }\right>_{\rm div}$ 
express the divergence part and 
is namely the DeWitt-Schwinger counter-term, 
which can be generally given by~\cite{Christensen:1976vb}
\begin{align}
&\left< {  \delta \phi \left(x\right) }{  \delta \phi \left(x'\right) }\right>_{\rm div} = \frac { 1 }{ 8{ \pi  }^{ 2 }\sigma  } +
\frac { { m }^{ 2 }+\left( \xi -1/6 \right) R }{ 8{ \pi  }^{ 2 } } \nonumber  \\ &
\times \left[ \gamma +
\frac { 1 }{ 2 } \ln { \left( \frac { { \mu  }^{ 2 }\sigma  }{ 2 }  \right)  }  \right] -\frac { { m }^{ 2 } }{ 16{ \pi  }^{ 2 } } 
+\frac { 1 }{ 96{ \pi  }^{ 2 } } { R }_{ \alpha \beta  }\frac { { \sigma  }^{ ;\alpha  }{ \sigma  }^{ ;\beta  } }{ \sigma  } ,
\end{align}
where $\sigma$ is the biscalar associated with the short geodesic,
$R$ or ${ R }_{ \alpha \beta  }$ are respectively the Ricci scalar or tensor and
$ \gamma$ express the Euler-Mascheroni constant.
The renormalization parameter $\mu$ corresponds 
to the mass $m$ of the scalar field and the massless case lead to the well-known ambiguity~\cite{Anderson:1994hg}, but the renormalization procedure can  
eliminate this ambiguity for $\left< {  T  }_{\mu \nu} \right>$ by the cosmological experiment or observation.
In the Schwarzschild metric for the massless scalar field where $m=0$ and $R=0$, 
we can simplify the DeWitt-Schwinger counter-term of 
$\left< {  \delta \phi \left(x\right) }{  \delta \phi \left(x'\right) }\right>_{\rm div}$ to be,
\begin{align}
\left< {  \delta \phi \left(x\right) }{  \delta \phi \left(x'\right) }\right>_{\rm div} = \frac { 1 }{ 8{ \pi  }^{ 2 }\sigma  } .
\end{align}

For simplicity we take the time separation $\epsilon$ between $x=\left( t,r,\theta,\varphi \right) $ and $x'=\left( t+\epsilon, r,\theta,\varphi \right) $
and the renormalized expression of  $\left< {  \delta \phi  }^{ 2 } \right>$ 
in the Boulware vacuum $\left|{ 0_{\rm B} } \right>$ is given by
\begin{align}
\left< {  \delta \phi  }^{ 2 }\left(x\right) \right>_{\rm ren}= & \lim _{ \epsilon \rightarrow 0} \Biggl[
\frac{1}{16\pi^2}{ \int _{ 0 }^{ \infty  }{ \frac { { e }^{ -i\omega \epsilon  }}{\omega  }  }  } 
\biggl[ \sum _{ l=0 }^{  \infty }\left(2l+1\right)
\left| { R }_{ l }^{in} \left( r;\omega  \right) \right|^{2}\nonumber \\ &+
\left| { R }_{ l }^{out} \left( r;\omega  \right)\right|^{2}\biggr]d\omega 
-\frac { 1 }{ 8{ \pi  }^{ 2 }\sigma \left( \epsilon  \right) }  \Biggr].  \label{eq:asfkdjj}
\end{align}
Taking a second-order geodesic expansion we get
the following expression~\cite{Christensen:1976vb}
\begin{align}
\sigma \left( \epsilon  \right) =-\frac { 1-2M_{\rm BH}/r }{ 2 } { \epsilon  }^{ 2 }
-\frac { { M }_{\rm BH}^{ 2 }\left( 1-2M_{\rm BH}/r \right)  }{ 24{ r }^{ 4 } } { \epsilon  }^{ 4 }+O\left( { \epsilon  }^{ 5 } \right) ,\label{eq:assdjj}
\end{align}
where $\epsilon^{-2}$ satisfy the following relation
\begin{align}
\epsilon^{-2} =- \int _{ 0 }^{ \infty  }{ \omega { e }^{ i\omega \epsilon  }d\omega  }. \label{eq:asdkj}
\end{align}
By using Eqs.~(\ref{eq:asfkdjj}), (\ref{eq:assdjj}) and (\ref{eq:asdkj}), 
we obtain the renormalized expression of 
the Boulware vacuum $\left|{ 0_{\rm B} } \right>$ as follows~\cite{Candelas:1980zt},
\begin{align}
&\left< { 0_{\rm B} }|{ \delta { \phi  }^{ 2 }\left( x \right)  }|{ 0_{\rm B} } \right>_{\rm ren}= 
\frac{1}{16\pi^2}{ \int _{ 0 }^{ \infty  }{ \frac { d\omega  }{ \omega  }  }  } \\ 
&\Biggl[ \sum _{ l=0 }^{  \infty }\left(2l+1\right)\Bigl[
\left| { R }_{ l }^{in}\left( r;\omega  \right) \right|^{2}+
\left| { R }_{ l }^{out} \left( r;\omega  \right)\right|^{2}\Bigr] \nonumber 
-\frac{4\omega^{2}}{1-2M_{\rm BH}/r}\Biggr]\\ &-\frac{M_{\rm BH}^{2}}{48\pi^{2}r^{4}\left(1-2M_{\rm BH}/r\right)}. \nonumber 
\end{align}

For the Boulware vacuum $\left|{ 0_{\rm B} } \right>$ we have the asymptotic expression 
of the renormalized vacuum fluctuations $\left< {  \delta \phi  }^{ 2 } \right>_{\rm ren}$~\cite{Candelas:1980zt},
\begin{align*}
&\left< { 0_{\rm B} }|{ \delta { \phi  }^{ 2 }\left( x \right)  }|{ 0_{\rm B} } \right>_{\rm ren}\longrightarrow \infty \quad \left(r\rightarrow 2M_{\rm BH}\right),\\
&\left< { 0_{\rm B} }|{ \delta { \phi  }^{ 2 }\left( x \right)  }|{ 0_{\rm B} } \right>_{\rm ren}\longrightarrow 1/r^{2} \quad \left( r\rightarrow \infty \right),
\end{align*} 
where the renormalized expression 
$\left< {  \delta \phi  }^{ 2 } \right>_{\rm ren}$ is singular on the event horizons 
$ r = 2M_{\rm BH}$ and ill-defined around the black-hole horizon. 
In the case of the energy momentum tensor $\left< {  T  }_{\mu \nu} \right>$,
the renormalized expression of the energy momentum tensor 
$\left< {  T  }_{\mu \nu} \right>_{\rm ren}$ has been 
given by Refs.\cite{Candelas:1980zt,Page:1982fm,Brown:1986jy,
Frolov:1987gw,Vaz:1988gh,Barrios:1990vg,Anderson:1994hg} and 
shows similar properties to the renormalized 
expression of $\left< {  \delta \phi  }^{ 2 } \right>_{\rm ren}$ 
%%%%%%%%%%%%%%%%%%%%%%%%%%%%%
%%%%%%%%%%%%%%%%%%%%%%%%%%%%%
\footnote{For the Boulware vacuum $\left|{ 0_{\rm B} } \right>$ 
the renormalized energy momentum tensor 
$\left< {  T  }_{\mu \nu} \right>_{\rm ren}$ is given by
\begin{align*}
&\left< { 0_{\rm B} }|{  T  }_{\mu}^{\nu }|{ 0_{\rm B} } \right>_{\rm ren}\longrightarrow 
-\frac{1}{30\cdot 2^{12} \pi^{2}M_{\rm BH}^{4}\left(1-2M_{\rm BH}/r\right)^{2}}  \\ 
& \quad\quad\quad\quad \times \begin{pmatrix} -1 & 0 & 0 & 0 \\ 0 & 1/3 & 0 & 0 \\ 0 & 0 & 1/3 & 0 \\ 0 & 0 & 0 & 1/3 \end{pmatrix} 
\quad \left(r\rightarrow 2M_{\rm BH}\right), \\
&\left< { 0_{\rm B} }|{  T  }_{\mu}^{\nu }|{ 0_{\rm B} } \right>_{\rm ren}\longrightarrow 
1/r^{6}\quad \left(r\rightarrow \infty \right),
\end{align*}
where the renormalized expression of $\left< {  T  }_{\mu \nu} \right>_{\rm ren}$ for $\left|{ 0_{\rm B} } \right>$ 
produces a negative energy density divergence at the black-hole horizon $ r = 2M_{\rm BH}$.
This fact originates from the infinite blueshift of the negative energy.}.
%%%%%%%%%%%%%%%%%%%%%%%%%%%%%
%%%%%%%%%%%%%%%%%%%%%%%%%%%%%
This state closely reproduces the Minkowski vacuum $\left|{ 0_{\rm M} } \right>$ at infinity 
because $\left< { 0_{\rm B} }\right|  {  \delta \phi  }^{ 2 }  \left|{ 0_{\rm B} } \right> \rightarrow 1/r^{2}$ in the limit 
$r \rightarrow \infty$. However, the Boulware vacuum is singular on the event horizon $r=2M_{\rm BH}$ 
and hence unacceptable near the black-hole horizon.
Thus, the usual interpretation of the above result is that 
the Boulware vacuum $\left|{ 0_{\rm B} } \right>$ 
is considered to be the appropriate vacuum state around a static star and not a black hole.

Next, let us consider the Unruh vacuum $\left|{ 0_{\rm U} } \right>$ which 
is formally defined by taking ingoing modes to be 
positive frequency with respect to $\partial_{t}$, but outgoing modes to be positive frequency 
with respect to the Kruskal coordinate $\partial_{U}$~\cite{Unruh:1976db}.
For the Unruh vacuum we obtain the two-point correlation function
$\left< {  \delta \phi  }^{ 2 } \right>$~\cite{Christensen:1977jc,Candelas:1980zt},
\begin{align}
&\left< { 0_{\rm U} }|{ \delta { \phi  }^{ 2 }\left( x \right)  }|{ 0_{\rm U} } \right>= 
\frac{1}{16\pi^2}{ \int _{ 0 }^{ \infty  }{ \frac { d\omega  }{ \omega  }  }  } \\ 
&\left[ \sum _{ l=0 }^{  \infty }\left(2l+1\right)\Bigl[\left| { R }_{ l }^{in} \left( r;\omega  \right) \right|^{2}+
\coth { \left( \frac { \pi \omega  }{ \kappa  }  \right)  }\left| { R }_{ l }^{out} \left( r;\omega  \right)\right|^{2}\Bigr] \right], \nonumber 
\end{align}
where we introduce $\kappa= (4M_{\rm BH})^{-1}$ which is the surface gravity of the black hole and 
the factor of $\coth { \left( \frac { \pi \omega  }{ \kappa  }  \right)  }$ originates from 
the thermal features of the outgoing modes.
The renormalized vacuum fluctuations in the Unruh vacuum $\left|{ 0_{\rm U} } \right>$ 
are given by 
\begin{align}
&\left< { 0_{\rm U} }|{ \delta { \phi  }^{ 2 }\left( x \right)  }|{ 0_{\rm U} } \right>_{\rm ren}= 
\frac{1}{16\pi^2}{ \int _{ 0 }^{ \infty  }{ \frac { d\omega  }{ \omega  }  }  }\label{eq:fvacuumg} \\ &\Biggl[ \sum _{ l=0 }^{  \infty }\left(2l+1\right)\Bigl[
\left| { R }_{ l }^{in} \left( r;\omega  \right) \right|^{2}+
\coth { \left( \frac { \pi \omega  }{ \kappa  }  \right)  }\left| { R }_{ l }^{out} \left( r;\omega  \right)\right|^{2}\Bigr] \nonumber 
\\ &-\frac{4\omega^{2}}{1-2M_{\rm BH}/r}\Biggr]-\frac{M_{\rm BH}^{2}}{48\pi^{2}r^{4}\left(1-2M_{\rm BH}/r\right)}. \nonumber 
\end{align}
For the Unruh vacuum $\left|{ 0_{\rm U} } \right>$, we can get an
asymptotic expression of the renormalized vacuum fluctuations,
\begin{align*}
&\left< { 0_{\rm U} }|{ \delta { \phi  }^{ 2 }\left( x \right)  }|{ 0_{\rm U} } \right>_{\rm ren}\longrightarrow \frac{1}{192\pi^{2}M_{\rm BH}^{2}} \\ &-
\frac{1}{32\pi^{2}M_{\rm BH}^{2}}\int _{ 0 }^{ \infty  }
{ \frac { d\omega \omega \sum _{ l=0 }^{ \infty  }{ \left( 2l+1 \right)  } { \left| { { B }_{ l } }\left( \omega  \right)  \right|  }^{ 2 } }
{ \omega\left({ e }^{ 2\pi \omega /\kappa  }-1 \right)}  }\ \left( r\rightarrow 2M_{\rm BH}\right),
\nonumber \\
&\left< { 0_{\rm U} }|{ \delta { \phi  }^{ 2 }\left( x \right)  }|{ 0_{\rm U} } \right>_{\rm ren}\longrightarrow 1/r^{2} \quad \left(r\rightarrow \infty \right).
\end{align*}
which corresponds to the evaporating black hole.
The Unruh vacuum $\left|{ 0_{\rm U}} \right>$ corresponds to the state where
the black hole radiates at the Hawking temperature $T_{\rm H}=1/(8\pi M_{\rm BH})$ in the empty space, and therefore, the vacuum fluctuations $\left< {  \delta \phi  }^{ 2 }  \right>$ 
approache the thermal fluctuations near the black-hole horizon to be
$\left< { 0_{\rm U} }\right|  {  \delta \phi  }^{ 2 }   \left|{ 0_{\rm U} } \right>  
\rightarrow \mathcal{O}\left(T_{\rm H}^{2}\right)$ in the limit $r \rightarrow 2M_{\rm BH}$.
Thus, the Unruh vacuum is considered to be an appropriate vacuum which 
describe the evaporating black hole formed by gravitational collapse.

Finally, we discuss the Hartle-Hawking Vacuum $\left|{ 0_{\rm HH}} \right>$ 
which is formally defined by taking ingoing modes to be positive frequency 
with respect to $\partial_V$, and outgoing modes to be positive frequency with
respect to the Kruskal coordinate $\partial_U$~\cite{Hartle:1976tp}.
In the Hartle-Hawking vacuum $\left|{ 0_{\rm HH} } \right>$, we can get,
\begin{align}
&\left< { 0_{\rm HH} }|{ \delta { \phi  }^{ 2 }\left( x \right)  }|{ 0_{\rm HH} } \right>= 
\frac{1}{16\pi^2}{ \int _{ 0 }^{ \infty  }{ \frac { d\omega  }{ \omega  }  }  } \\ 
&\left[ \coth { \left( \frac { \pi \omega  }{ \kappa  }  \right)  } \sum _{ l=0 }^{  \infty }\left(2l+1\right)\Bigl[
\left| { R }_{ l }^{in}  \left( r;\omega  \right) \right|^{2}+
\left| { R }_{ l }^{out}  \left( r;\omega  \right)\right|^{2}\Bigr] \right] \nonumber .
\end{align}
For $\left|{ 0_{\rm HH} } \right>$, we obtain the asymptotic expression of
the renormalized vacuum fluctuations,
\begin{align*}
&\left< { 0_{\rm HH} }|{ \delta { \phi  }^{ 2 }\left( x \right)  }|{ 0_{\rm HH} } \right>_{\rm ren}
\longrightarrow \frac{1}{192\pi^{2}M_{\rm BH}^{2}}\quad   \left(r\rightarrow 2M_{\rm BH}\right), \\
&\left< { 0_{\rm HH} }|{ \delta { \phi  }^{ 2 }\left( x \right)  }|{ 0_{\rm HH} } \right>_{\rm ren}\longrightarrow T^{2}_{\rm H}/12
\quad \left(r\rightarrow \infty \right),
\end{align*}
where $\left< {  \delta \phi  }^{ 2 }  \right>_{\rm ren}$ 
becomes exactly the thermal fluctuation at infinity, i.e 
$\left< { 0_{\rm HH} }\right|  {  \delta \phi  }^{ 2 }   \left|{ 0_{\rm HH} } \right>  
\rightarrow T_{\rm H}^{2}/12$ in the limit $r \rightarrow \infty $.
Thus, the Hartle-Hawking vacuum corresponds to a black hole 
in thermal equilibrium at $T_{\rm H}$.
If we consider the vacuum fluctuations around the black hole 
without no thermal radiation, i.e. the temperature of the universe is  
lower than the Hawking temperature $T_{\rm H}$, this vacuum state 
is not adequate to describe the evaporation of the black hole.

The analytic approximations of
$\left< { \delta \phi }^{ 2 } \right>_{\rm ren}$ and
$\left< { T }_{\mu \nu} \right>_{\rm ren}$ in Schwarzschild
spacetime for the various vacua (Boulware, Unruh vacuum and
Hartle-Hawking) and the various fields of spin $0$, $1/2$ and $1$
has been studied by Ref.\cite{Candelas:1980zt,Page:1982fm,Howard:1984qp,
Howard:1985yg,Brown:1986jy,Frolov:1987gw,Vaz:1988gh,Barrios:1990vg,
Nugaev:1991aa,Anderson:1994hg,Matyjasek:1996dm,
Matyjasek:1996if,Matyjasek:1996ig,Matyjasek:1996ih,Visser:1997sd,Matyjasek:1998zs,Matyjasek:1998mq}.
The renormalized expression of the various fields are proportional to
the inverse of the black-hole mass $M_{\rm BH}$ near the black-hole
horizon and approximately approach the thermal fluctuations with the
Hawking temperature $T_{\rm H}$ besides the Boulware vacuum.
The quantum effects of the vacuum fluctuations expressed by
$\left< { \delta \phi }^{ 2 } \right>_{\rm ren}$ and
$\left< { T }_{\mu \nu} \right>_{\rm ren}$ determine 
the vacuum stability and the evaporation of the black hole.
We recall that the Unruh vacuum is an appropriate vacuum state 
describing the vacuum fluctuations around evaporating black holes.
Thus, it is reasonable to assume that the vacuum fluctuations of the 
various field like the Higgs, $W$ and $Z$ bosons and the top quark 
approximately approach the Hawking thermal fluctuations 
around the black hole horizon,
\begin{align}
\left< {  \delta \phi  }^{ 2 } \right>_{\rm ren}&
\approx \left< {  \delta W }^{ 2 } \right>_{\rm ren}
\approx \left< {  \delta Z}^{ 2 } \right>_{\rm ren}
\approx \left< {  \delta t }^{ 2 } \right>_{\rm ren}\\
&\approx \frac{T_{\rm H}^2}{3} -
2T_{\rm H}^2\int _{ 0 }^{ \infty  }
{ \frac { d\omega \omega \sum _{ l=0 }^{ \infty  }{ \left( 2l+1 \right)  } 
{ \left| { { B }_{ l } }\left( \omega  \right)  \right|  }^{ 2 } }
{ \omega\left({ e }^{ 2\pi \omega /\kappa  }-1 \right)}  }.\nonumber 
\end{align}

%%%%%%%%%%%%%%%%%%%%%%%%%%%%%%%%%%%%%%
%%%%%%%%%%%%%%%%%%%%%%%%%%%%%%%%%%%%%%
\section{The Higgs potential around evaporating block holes}
\label{sec:block holes}
%%%%%%%%%%%%%%%%%%%%%%%%%%%%%%%%%%%%%%
%%%%%%%%%%%%%%%%%%%%%%%%%%%%%%%%%%%%%%
In this section we consider the standard model (SM)
effective Higgs potential $V_{\rm eff}\left( \phi  \right)$
around evaporating block holes
where $\phi$ is the Higgs field.
The one-loop Higgs potential in 't Hooft-Landau gauge and 
$\overline {\rm  MS } $ scheme
is given by~\cite{Ford:1992mv,Casas:1994qy,Espinosa:1995se},
\begin{align}
V_{\rm eff}&\left( \phi  \right)
=\rho_{\Lambda}( \mu )+\frac{m^{2}_{ \phi }( \mu )}{2}\phi^{2}
+\frac{\lambda_{ \phi }(\mu )}{4}\phi^{4} \nonumber \\  &
+\sum _{  i=W,Z,t,G,H }{ \frac { { n }_{ i } { M }_{ i }^{ 4 }
\left( \phi  \right)  }{ 64{ \pi  }^{ 2 } }
\left[ \log{ \frac { { M }_{ i }^{ 2 }\left( \phi  \right)  }
{ \mu^{ 2 } }  } -{ C }_{ i } \right]  } \label{eq:fhlgkdg},
\end{align}
The coefficients $n_{i}$ and $C_{i}$ are given by
\begin{align*}
&{ n }_{ W }=6,\ { n }_{ Z }=3,\ { n }_{ t }=-12,\ { n }_{ G }=3,\ { n }_{ H }=1, \nonumber \\
&{ C }_{ W }={ C }_{ Z }=5/6,\ { C }_{ t }={ C }_{ G }={ C }_{ H }=3/2,
\end{align*}
and the mass terms ${ m }_{ i }^{ 2 }\left( \phi  \right)$ of the $W$ and $Z$ bosons, the top quark, 
the Nambu-Goldstone bosons, and the Higgs boson are give by
\begin{align*}
&{ m }_{ W }^{2}=\frac{1}{4}g^{2}\phi^{2}, \  { m }_{ Z }^{2}=\frac{1}{4}\left[ g^{2} + g'^{2}\right] \phi^{2},\ { m }_{ t }^{2}=\frac{1}{2}y^{2}_{t}\phi^{2}, \\
&{ m }_{ G }^{2}={ m }_{ \phi }^{2}+\lambda_{ \phi }\phi^{2} ,\ { m }_{ H }^{2}={ m }_{\phi }^{2}+3\lambda_{ \phi }\phi^{2} ,
\end{align*}
where $g$, $g'$, $y_{t}$ are the $SU(2)_{L}$, $U(1)_{Y}$, top Yukawa couplings
and $\lambda_{ \phi }$ is the Higgs self-coupling. 
The potential of Eq.~(\ref{eq:fhlgkdg}) express the 
Higgs effective potential in flat spacetime, not curved spacetime.
Actually, we must include the backreaction from
the vacuum fluctuations of the various fields into the Higgs potential.
The modified effective Higgs potential including the 
Higgs, $W$ and $Z$ bosons and top quark vacuum fluctuations
can be given by~\cite{Kohri:2017iyl}:
\begin{align}
\begin{split}
&V_{\rm eff}\left( \phi  \right)=\rho_{\Lambda}( \mu )+\frac{m^{2}_{ \phi }( \mu )}{2}\phi^{2}
+\frac{\lambda_{ \phi }(\mu )}{4}\phi^{4}   \\
&+\frac{3\lambda(\mu)}{2}\left< {\delta \phi  }^{ 2 } \right>_{\rm ren}\phi^{2}
+\frac{g^{2}(\mu)}{8}\left< {  \delta W }^{ 2 } \right>_{\rm ren}\phi^{2}  \\  
& + \frac{\left[ g^{2}(\mu) + g'^{2}(\mu)\right]}{8}
\left< {  \delta Z }^{ 2 } \right>_{\rm ren}\phi^{2} 
  +\frac{y^{2}_{t}(\mu)}{4}\left< {  \delta t }^{ 2 } \right>_{\rm ren}\phi^{2}  \\  
& +\sum _{  i=W,Z,t,G,H }{ \frac { { n }_{ i } { M }_{ i }^{ 4 }\left( \phi  \right)  }{ 64{ \pi  }^{ 2 } } \left[ \log{ \frac { { M }_{ i }^{ 2 }\left( \phi  \right)  }{ \mu^{ 2 } }  } -{ C }_{ i } \right]  } \label{eq:fhlgkdg},
\end{split}
\end{align}
The Higgs vacuum fluctuations induce the false vacuum decay but 
the $W$ and $Z$ bosons and the top quark fluctuations stabilize 
the Higgs potential via the interaction.
Now let us rewrite the effective Higgs potential of Eq.~(\ref{eq:fhlgkdg})
by using the Unruh vacuum $\left|{ 0_{\rm U} } \right>$.
The renormalized vacuum fluctuations are 
proportional to the inverse of the black-hole mass $M_{\rm BH}$ and
the Hawking temperature $T_{\rm H}$.
Thus, it is now evident that the effective Higgs potential 
around evaporating black holes
reproduce the thermal case~\cite{Arnold:1991cv,Carrington:1991hz,
Anderson:1991zb,Delaunay:2007wb,Rose:2015lna} and generally 
is stabilized.
The Higgs potential $V_{\rm eff}\left( \phi  \right) $ 
including the vacuum fluctuations of the various fields around 
evaporating black holes can be approximately written as,
\begin{align}
V_{\rm eff}\left( \phi  \right)   \simeq \mathcal{O}
\left( T_{\rm H}^{2} \right)\phi^{2}
+\frac{\lambda_{ \rm eff}(\phi)}{4}\phi^{4} ,\label{eq:thamlhiggs}
\end{align}
where $\lambda_{ \rm eff}(\phi)$ is the effective Higgs 
self-coupling defined by the effective potential $V_{\rm eff}\left( \phi  \right) $ 
and the maximal field value ${  \phi  }_{ \rm max }$ 
can be given by~\cite{Kohri:2016wof},
\begin{align}
{  \phi  }_{ \rm max } \approx \mathcal{O}\left(10\right)\sqrt{
V_{\rm eff}''\left( \phi  \right)}\approx\mathcal{O}
\left(1\sim10 \right)\cdot T_{\rm H}\,  \gtrsim \, \Lambda_{I} \label{eq:thaml}
\end{align}
Thus, it turns out that the Hawking radiation generally stabilizes the 
Higgs potential but it does not mean that the Higgs vacuum is perfectly stable 
around evaporating black holes. In the next section we discuss the 
the Higgs vacuum stability around the black hole in more detail.

%%%%%%%%%%%%%%%%%%%%%%%%%%%%%%%
%%%%%%%%%%%%%%%%%%%%%%%%%%%%%%%
\section{The false vacuum decay around evaporating block holes}
\label{sec:electroweak}
%%%%%%%%%%%%%%%%%%%%%%%%%%%%%%%
%%%%%%%%%%%%%%%%%%%%%%%%%%%%%%%
The false vacuum decay is usually analyzed by
the Euclidean or bounce solution where one idealize the system that 
a bubble of true vacuum is surround by the potential wall 
and false vacuum sea.
The created vacuum bubble expands eating other regions of 
false vacuum and a first order phase transition is complete.
The vacuum decay ratio and 
critical bubble size can be determined by  
the Euclidean solution or the instanton which capture 
the catalysis induced by quantum fluctuations 
around the vacuum~\cite{Coleman:1977py,Callan:1977pt}.
The analysis of the false vacuum decay in flat spacetime is 
completely understood, but it is not in curved spacetime.
Taking into account quantum effects of the gravity, 
induced vacuum fluctuations and gravitational particle production 
occurs, and the process of the false vacuum decay is 
expected to be different from what is considered in flat spacetime.
In~\cite{Coleman:1980aw}, Coleman and de Luccia (CDL) 
formulate the approach of the 
false vacuum decay based on the instanton.
The CDL instanton simply describes a bubble nucleation in 
curved spacetime.
The decay rate of the vacuum in curved spacetime is given by
\begin{align}
\Gamma_{\rm decay}\left(  \phi \right) = A\exp(-B),
\end{align}
where $A$ is a prefactor and $B$ is given by the action of the bounce solution. 
Following~\cite{Gregory:2013hja,Burda:2015isa,Burda:2015yfa,Burda:2016mou},
we review the false vacuum decay around the black hole.
For simplicity, we introduce 
two Schwarzschild geometry with different cosmological constants 
separated by a thin wall tension, 
\begin{align}
ds^2&=f(r)d\tau_\pm^2+{dr^2\over f(r)}+r^2d\Omega^2,\nonumber \\
f(r)&\equiv 1-\frac{2GM_\pm}{r}-\frac{\Lambda_\pm r^2}{3}, 
\end{align}
where $\tau_\pm$ are different time coordinates,
$M_\pm$ are different black hole masses and 
$\Lambda_\pm$ are different cosmological constants
where we simply assume that Higgs false vacuum has 
$\Lambda_+=0$, and true vacuum has $\Lambda_- = -3/\ell^2$.
In this geometry, the bounce action with the black hole
can be given by~\cite{Gregory:2013hja},
\begin{align}
&B={{\cal A}_h^+\over 4G}-{{\cal A}_h^-\over 4G}
\label{mplusminusaction}\\
&+{1\over 4G}\oint d\lambda \left\{\left ( 2R - 6GM_+\right)\dot{\tau}_+
- \left (2R-6GM_-\right) \dot{\tau}_-\right\}\nonumber
\end{align}
where $R(\tau_\pm)$ is the bubble radius
and ${\cal A}_h^\pm$ are the black hole horizon areas. 
The prefactor $A$ can be estimated by
this bounce action $B$ and the vacuum decay ratio 
including the black hole can be roughly rewritten as~\cite{Burda:2015isa},
\begin{equation}
\Gamma_{\rm decay}\left(  \phi \right)
\approx \left({B\over 2\pi }\right)^{1/2}\left(GM_+\right)^{-1}e^{-B}.
\label{decay}
\end{equation}
where $B(M_+/M_{\rm P}, {  \phi  }_{ \rm max })$ depend on the black hole mass
$M_+$ and the maximal field value ${  \phi  }_{ \rm max }$ of the Higgs.
If the black hole mass approaches the Planck mass
with the Hawking temperature 
$T_{\rm H}\approx  M_{\rm P}\approx 10^{19}\ {\rm GeV}$
where $M_{\rm P}$ is the Planck mass,
a false vacuum decay around black hole is 
sufficiently enhanced~\cite{Burda:2015isa}.
However, the barrier of the Higgs potential is sufficiently 
lifted up by the Hawking radiation from Eq.~(\ref{eq:thaml}).
Taking into account the back-reaction effects of the 
Hawking radiation or the vacuum fluctuations 
around the black hole, the decay rate is exponentially suppressed.

Generally, the analysis of the Euclidean methods or the instanton
is standard to investigate the false vacuum decay. However, 
they are physically obscure and somewhat suspicious in curved spacetime.
Recalling that we have no unique vacua in 
curved spacetime (the Schwarzschild spacetime has three well defined vacua, 
Boulware, Unruh and Hartle-Hawking vacuum), different vacuum states 
lead to different vacuum fluctuations and different decay ratios.
Thus, one needs to understand what kind of process occurs in the 
vacuum collapse around the black hole and consider 
an another derivation of the vacuum decay rate 
without relying on the instanton.

Now, we provide another approach to investigate the false vacuum 
decay around evaporating black holes by 
using the renormalized two-point correlation function of Eq.~(\ref{eq:fvacuumg})
in Unruh vacuum. This approach using the two-point correlation
function $\left< { \delta \phi }^{ 2 } \right>$ 
is consistent with the instanton method~\cite{Linde:1991sk,
Linde:1978px,Dine:1992wr} and extremely simple even 
in the investigation of the false vacuum decay
with taking account of the gravitational effects.

Let us introduce the vacuum decay rate by using
the renormalized two-point correlation function.
The probability of the local Higgs fields where 
the vacuum fluctuations $\left< {  \delta \phi  }^{ 2 } \right>_{\rm ren}$ exist
can be given by~\cite{Espinosa:2015qea},
\begin{align}
P\left( \phi \right) 
= \frac { 1 }{ \sqrt {2{ \pi  }\left< {  \delta \phi  }^{ 2 } \right>_{\rm ren} } } 
\exp \left( -\frac {{  \phi  }^{ 2 } }
{ 2\left< {  \delta \phi  }^{ 2 } \right>_{\rm ren} }  \right)
\label{eq:hqqsdkfg}.
\end{align}
By using Eq.~(\ref{eq:hqqsdkfg}), 
we can obtain the probability not to exceed 
the hill of the Higgs potential as follows~\cite{Espinosa:2015qea}:
\begin{align}
{ P }\left(  \phi<{ \phi }_{ \rm max }\right) 
&\equiv \int _{ -{ \phi  }_{\rm  max } }^{ { \phi }_{ \rm max } }
{ P\left( \phi, \left< {  \delta \phi  }^{ 2 } \right>_{\rm ren} \right) d\phi }, 
\nonumber \\ &= {\rm erf}\left( \frac{{ \phi }_{ \rm max }  }
{ \sqrt { 2\left< {  \delta \phi  }^{ 2 } \right>_{\rm ren}  }}  \right)
\label{eq:hswegg},
\end{align}
where ${ \phi }_{ \rm max }$ is the maximal field value
of the Higgs potential. Considering the probability that the localized
Higgs fields go into the negative true vacuum, the vacuum decay ratio 
is estimated to be
\begin{align}
\Gamma_{\rm decay}\left(  \phi \right)
&\equiv{ P }\left(  \phi>{ \phi }_{ \rm max }\right)=
1- {\rm erf}\left( \frac{{ \phi }_{ \rm max }  }
{ \sqrt { 2\left< {  \delta \phi  }^{ 2 } \right>_{\rm ren}  }}  \right),\nonumber
\\ &\simeq 
\frac {\sqrt{2\left< {  \delta \phi  }^{ 2 } \right>_{\rm ren}  }}{\pi{ \phi }_{ \rm max }}
\exp \left( -\frac {{  \phi  }^{ 2 }_{ \rm max }  }{ 2\left< {  \delta \phi  }^{ 2 } \right>_{\rm ren} }  \right) \label{eq:afklsjiijgg} .
\end{align}
Then, the constraint from the vacuum decay for the Higgs field is represented by
%%%%%%%%%%%%%%%%%%%%%%%%%%%%%
%%%%%%%%%%%%%%%%%%%%%%%%%%%%%
\footnote{The detail analysis of the Higgs vacuum decay induced 
by thermal fluctuations was investigated by~\cite{Kohri:2016wof}.}
%%%%%%%%%%%%%%%%%%%%%%%%%%%%%
%%%%%%%%%%%%%%%%%%%%%%%%%%%%%
\begin{align}
\mathcal{N}_{\rm PBH}\cdot\Gamma_{\rm decay}\left(  \phi \right)
\lesssim1,\label{eq:aaawegg}
\end{align}
where $\mathcal{N}_{\rm PBH}$ is the number of the evaporating (or evaporated) primordial black holes during the cosmological history of the Universe.
Substituting Eq.~(\ref{eq:afklsjiijgg}) into
Eq.~(\ref{eq:aaawegg}), we can simplify the constraint of the
vacuum stability,
\begin{align}
\frac{\left< {  \delta \phi  }^{ 2 } \right>_{\rm ren}}{{ \phi }_{ \rm max }^{2}} \lesssim\frac{1}{ 2}\left(\log {\mathcal{N}_{\rm PBH}  }\right)^{-1}
\label{eq:hsdedg},
\end{align}
in order not to induce a vacuum transition due to large Higgs fluctuations.
In the Unruh vacuum $\left|{ 0_{\rm U} } \right>$ 
corresponding to the vacuum state 
around evaporating black holes, 
$\left< {  \delta \phi  }^{ 2 } \right>_{\rm ren} $
approximately take the value of
the Hawking temperature $T_{\rm H}$ near the horizon.
However, at the infinity $\left< {  \delta \phi  }^{ 2 } \right>_{\rm ren} $ 
attenuates rapidly and becomes zero.
Thus, we ontain the renormalized vacuum 
fluctuations $\left< {  \delta \phi  }^{ 2 } \right>_{\rm ren} $ 
around evaporating black holes as follows:
\begin{align}
\left< {  \delta \phi  }^{ 2 } \right>_{\rm ren}\simeq \mathcal{O}\left(10^{-2}\sim
10^{-1} \right)\cdot T_{\rm H}
\quad \left(r\rightarrow 2M_{\rm BH}\right)\label{eq:kkjdfjds}.
\end{align}
However, the most uncertain thing in the stochastic formalism is 
how to determine the volume factor of $\mathcal{V}$. We took $\mathcal{V}$ to be the volume of  the domains in which
the vacuum fluctuation $\left< {  \delta \phi  }^{ 2 } \right>_{\rm ren} $ governs.
By using Eqs.~(\ref{eq:thaml}), (\ref{eq:hsdedg}) and (\ref{eq:kkjdfjds}),
we can estimate a constraint on the number of the evaporating primordial black holes as follows
\begin{align}
\mathcal{N}_{\rm PBH}\cdot\Gamma_{\rm decay}\left(  \phi \right) & \simeq 
\frac {\mathcal{N}_{\rm PBH}\sqrt{2\left< {  \delta \phi  }^{ 2 } \right>_{\rm ren}  }}{\pi{ \phi }_{ \rm max }}
\exp \left( -\frac {{  \phi  }^{ 2 }_{ \rm max }  }{ 2\left< {  \delta \phi  }^{ 2 } \right>_{\rm ren} }  \right) \nonumber \\
& \approx {\mathcal{N}_{\rm PBH}}\cdot  
e^{-\mathcal{O}\left(10^{\, 2\sim3} \right)}\lesssim \ 1 \label{eq:afjgg},
\end{align}
where the left-hand side shows the number of the
primordial black holes ${\mathcal{N}_{\rm PBH}}$ within the presently visible part of the Universe
which cause the Higgs vacuum collapse.
Thus, we obtain a new bound on the number of the
evaporating primordial black holes to be
$ {\mathcal{N}_{\rm PBH}} \lesssim \mathcal{O}\left(10^{\,43\sim 434}
\right)$ which is extremely huge to 
threaten the Higgs metastability. 

Next, let us consider the upper bound on the yield of the PBHs
$Y_{ \rm PBH }/ \equiv { n }_{ \rm PBH }/s$ as follows,
\begin{eqnarray}
  \label{eq:bound0}
Y_{\rm PBH}=\frac { { n }_{ \rm PBH } }{ s } =
\frac { \mathcal{N}_{\rm PBH}}{ s_{ 0 }/{ H }_{ 0 }^{ 3 } } \lesssim 
\mathcal{O}\left(10^{-43}\right),
\end{eqnarray}
where $s_{0}$ denotes the entropy density at present
($\approx \left(3\times 10^{-4}\ {\rm eV}\right)^{3}$), and $H_{0}$ is
the current Hubble constant ($\approx 10^{-33}\ {\rm eV} )$. Note that
$Y_{\rm PBH}$ is constant from the formation time to the evaporation time.

It is convenient to transform this bound into an upper bound on
$\beta$, which is defined by taking
values at the formation of the PBH to be
\begin{eqnarray}
  \label{eq:beta0}
  \beta 
  &\equiv&
  \left.
  \frac{\rho_{\rm PBH}}{\rho_{\rm tot}}
  \right|_{\rm formation},
\end{eqnarray}
where $\rho_{\rm PBH}$ and $\rho_{\rm tot}$ are the energy density of
the PBHs and the total energy density of the Universe including the
PBHs at the formation, respectively. It is remarkable that $\beta$
means the number of the PBHs per the horizon volume at the formation
($\beta \sim n_{\rm PBH}/H^3$). Then we have a
relation~\cite{Carr:2009jm},
\begin{eqnarray}
  \label{eq:beta1}
  \beta 
   \sim
   10^{30} 
%  \left(
  \frac{n_{\rm PBH}}{s}
%  \right)
  \left(
  \frac{m_{\rm PBH}}{10^{15}{\rm g}}
  \right)^{3/2}.
\end{eqnarray}
Combining this relation with (\ref{eq:bound0}), we obtain
%%%%%%%%%%%%%%%%%%%%%%%%%%%%%
%%%%%%%%%%%%%%%%%%%%%%%%%%%%%
\footnote{The Hawking temperature is 
$T_{\rm H}=1.058\, (10^{13} {\rm g}/m_{\rm PBH})\ {\rm GeV}$.}
%%%%%%%%%%%%%%%%%%%%%%%%%%%%%
%%%%%%%%%%%%%%%%%%%%%%%%%%%%%
%%
\begin{eqnarray}
  \label{eq:PBHbound1}
  \beta  
   &\lesssim&
  \mathcal{O}\left(10^{-21} \right)
   \left(
  \frac{m_{\rm PBH}}{10^{9}{\rm g}}
  \right)^{3/2}.
\end{eqnarray}
This bound can be stronger than the known one for
$m_{\rm PBH} \lesssim 10^{9}{\rm g}$~\cite{Carr:2009jm}.
Fortheremore, at the final stage of the black hole evaporation, 
the black-hole mass $M_{\rm BH}$ becomes extremely small and 
the Hawking temperature $T_{\rm H}$ approaches to the Planck scale $M_{\rm P}\approx 10^{19}\ {\rm GeV}$.
Thus, the UV corrections of the beyond Standard Model (BSM) and 
the quantum gravity (QG) can not be ignored
and undoubtedly contribute to the false vacuum decay around the black hole.
When the Hawking temperature approaches to 
the Planck scale as $T_{\rm H} \rightarrow M_{\rm P}$,
the contributions of the BSM and the QG determine the stability of the vacuum. 
if these physics destabilize the
Higgs potential at the high energy, 
even a single evaporating black hole cause a vacuum decay of the Universe.

%%%%%%%%%%%%%%%%%%%%%%%%%%%%%
%%%%%%%%%%%%%%%%%%%%%%%%%%%%%
\begin{figure}[t]
\includegraphics[width=90
mm]{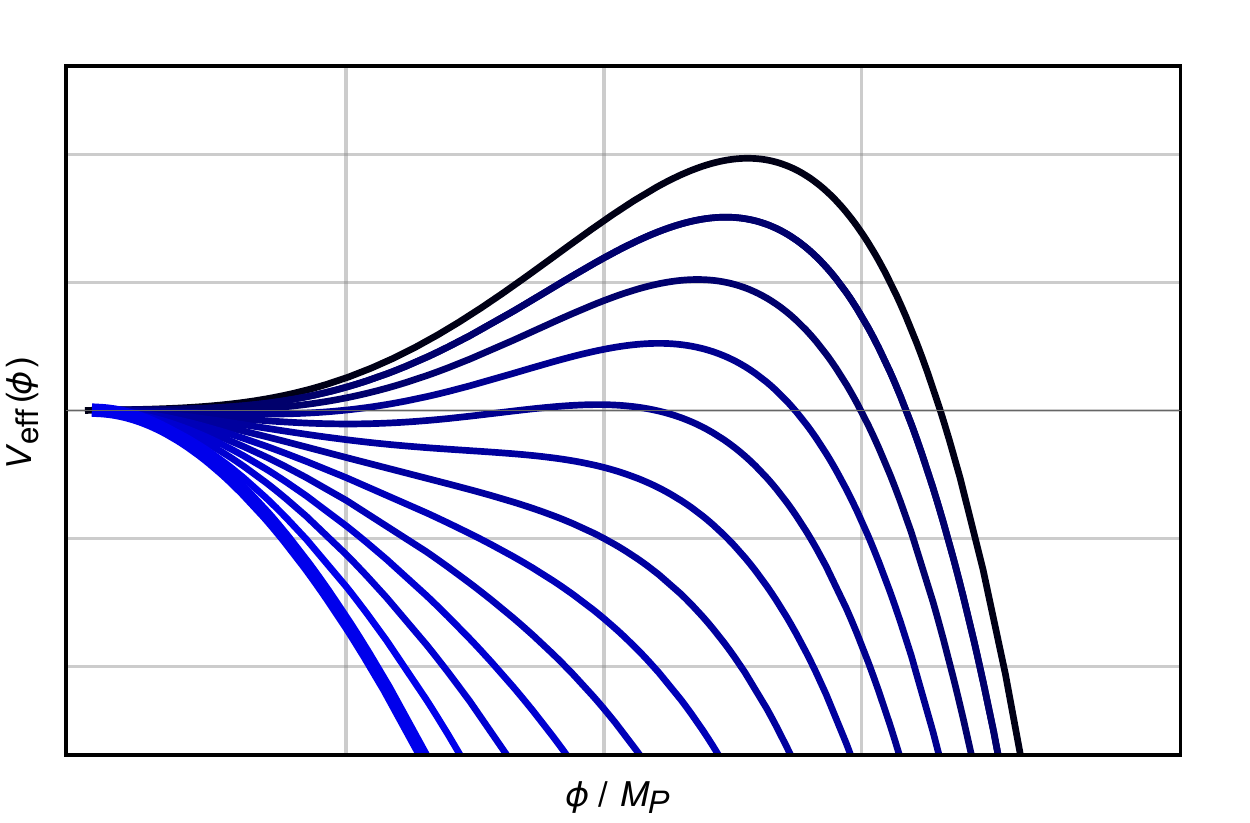}
\caption{The schematic diagram of the effective Higgs potential 
including higher-order corrections of $\lambda_{6}, \lambda_{8}<0$. 
The back-reaction of the Hawking radiation 
destabilizes the Higgs potential at $T_{\rm H}\gtrsim \Lambda_{\rm UV}$  
as shown by Eq.~(\ref{eq:uvfdf}) and the second-order transition 
occurs around the black hole.}
\label{Fig:qg1}
\end{figure}
%%%%%%%%%%%%%%%%%%%%%%%%%%%%%%
%%%%%%%%%%%%%%%%%%%%%%%%%%%%%%

Now, we consider the effective Higgs potential 
including corrections of the BSM and the QG.
For convenience we add two higher dimension operators $\phi^{6}$ and $\phi^{8}$ via the UV physics to the Higgs potential as follows,
\begin{align}
V_{\rm eff}\left( \phi  \right) =\frac{\lambda_{ \rm eff}(\phi)}{4}\phi^{4}+\frac{ \lambda_{ 6}}{6}\frac{\phi^{6}}{ \Lambda_{\rm UV}^{2} }+\frac{\lambda_{8}}{8}\frac{\phi^{8}}{ \Lambda_{\rm UV}^{4} }+\cdots
\label{eq:fdfkdkfdf},
\end{align}
where $\lambda_{6}$ and $\lambda_{8}$ are dimensionless coupling constants.
These contributions of $\lambda_{6}$ and $\lambda_{8}$ 
are usually negligible except for 
the large field excursion $\phi \approx \Lambda_{\rm UV}$
although they can affect the false vacuum decay 
through quantum tunneling~\cite{Branchina:2013jra,Lalak:2014qua,Branchina:2014usa,Branchina:2014rva}.
However, in the final evaporation of the black hole 
where $T_{\rm H} \rightarrow M_{\rm Pl}$, 
these quantum corrections of $\lambda_{6}$ and $\lambda_{8}$ 
can not be neglected 
and have a strong impact on the vacuum stability.
As previously discussed in Sec.~\ref{sec:electroweak}, 
the effective potential is modified 
by vacuum fluctuations around the black hole
and can be given as,
\begin{align}
\begin{split}
V_{\rm eff}\left( \phi  \right)  &\simeq \mathcal{O}
\left( T_{\rm H}^{2} \right)\phi^{2}+\frac{\lambda_{ \rm eff}(\phi)}{4}\phi^{4}\\
&+\frac{ \lambda_{ 6}\cdot \mathcal{O}\left( T_{\rm H}^{4} \right)}
{ 6\Lambda_{\rm UV}^{2} }\phi^{2} +\frac{ \lambda_{ 8}\cdot \mathcal{O}
\left( T_{\rm H}^{6} \right)}{ 8\Lambda_{\rm UV}^{4} }\phi^{2}+\cdots ,\label{eq:uvfdf}
\end{split}
\end{align}
which destabilizes at $T_{\rm H} \approx \Lambda_{\rm UV}$
when these dimensionless couplings are negative 
$\lambda_{6}, \lambda_{8}<0$. 
In this case the derivative of the Higgs potential 
becomes negative as
${\partial  V_{\rm eff}\left( \phi  \right)} / { \partial \phi  }< 0$, 
the local Higgs fields around the black hole classically roll down 
into the negative Planck-scale true vacuum (see Fig.~\ref{Fig:qg1})
and the Higgs Anti-de Sitter (AdS) bubbles 
whose sizes are about the black-hole horizon are formed.
Not all Higgs AdS bubbles threaten the Universe, 
which highly depends on their evolutions (see Ref.\cite{Espinosa:2015qea,Tetradis:2016vqb} for the detail discussions).
However, the Higgs AdS bubbles generally expand eating other regions of the false vacuum and finally consume the entire Universe. 
Thus, even a single evaporating black hole within the presently visible part of the Universe
can cause a vacuum decay of the Higgs 
at $T_{\rm H} \approx \Lambda_{\rm UV}$ although
this possibility strongly depends on the BSM or the QG
and the detail of the evaporation of the black hole.

\medskip
%%%%%%%%%%%%%%%%%%%%%%%%%%%%%%%%
%%%%%%%%%%%%%%%%%%%%%%%%%%%%%%%%%%%%%%
\section{Conclusion} 
\label{sec:conclusion}
%%%%%%%%%%%%%%%%%%%%%%%%%%%%%%
%%%%%%%%%%%%%%%%%%%%%%%%%%%%%%%%%%%%%%%%
In this paper, we have investigated the electroweak vacuum stability around 
evaporating black holes.
First, following~\cite{Candelas:1980zt} we have confirmed
that the Unruh vacuum $\left|{ 0_{\rm U} } \right>$ is an appreciate vacuum state 
to describe the evaporating black hole. Then, 
we have provide a new approach to investigate 
the false vacuum decay around the black hole 
using the renormalized vacuum fluctuations
$\left< { \delta \phi }^{ 2 } \right>_{\rm ren}$ in 
the Unruh vacuum $\left|{ 0_{\rm U} } \right>$. 
Clearly, we have shown
how vacuum fluctuations of the Higgs field
induce a collapse of the electroweak vacuum.

Furthermore, we have pointed out that back-reactions of the Hawking
thermal radiation can not be ignored and the Higgs potential is generally
stabilized by
the vacuum fluctuations around the black hole which was ignored in
previous works of the Higgs vacuum stability.
Incorporating the back-reaction effects and reanalyzing the
stability of the Higgs vacuum around the black hole, the stability
conditions approximately reproduce
the one for the thermal situation. 
Thus, one evaporating black hole
does not cause serious problems on the standard model Higgs
vacuum. However, a large number of the evaporating
(or evaporated) primordial black holes still threaten the vacuum
and we have obtained a new bound on the evaporating PBH abundance 
$\beta \lesssim
\mathcal{O}\left(10^{-21}\right) 
\left({m_{\rm PBH}}/{10^{9}{\rm g}}
\right)^{3/2}$ not to induce a collapse of the Universe.

Finally, we have discussed a possibility 
at the final stage of the evaporation of the black hole where the
black-hole mass $M_{\rm BH}$ becomes extremely small and the Hawking
temperature approaches the Planck scale $M_{\rm P}$.
Thus, the Planck scale physics or the beyond Standard Model (BSM)
directly may intervene and have a strong impact on the vacuum
stability around the black holes. Our discussion will be changed if the
Planck-scale physics or the BSM destabilize the Higgs potential.

\medskip
%%%%%%%%%%%%%%%%%%%%
\para{Acknowledgements} 
This work of K.K. is supported in part by MEXT KAKENHI
Nos.~JP15H05889, JP18H04594 and JP16H00877, and JSPS KAKENHI Nos. 26247042 and
JP1701131.
%%%%%%%%%%%%%%%%%%%%

%%%%%%%%%%%%%%%%%%%%%%%%%%%%%%%%%%%%%%%%%%%%%%%%%%%%%%%%%%%%%%%%%%%%
\nocite{}
\bibliography{Black-hole}

\providecommand{\noopsort}[1]{}\providecommand{\singleletter}[1]{#1}%
\begin{thebibliography}{100}%
\makeatletter
\providecommand \@ifxundefined [1]{%
 \@ifx{#1\undefined}
}%
\providecommand \@ifnum [1]{%
 \ifnum #1\expandafter \@firstoftwo
 \else \expandafter \@secondoftwo
 \fi
}%
\providecommand \@ifx [1]{%
 \ifx #1\expandafter \@firstoftwo
 \else \expandafter \@secondoftwo
 \fi
}%
\providecommand \natexlab [1]{#1}%
\providecommand \enquote  [1]{``#1''}%
\providecommand \bibnamefont  [1]{#1}%
\providecommand \bibfnamefont [1]{#1}%
\providecommand \citenamefont [1]{#1}%
\providecommand \href@noop [0]{\@secondoftwo}%
\providecommand \href [0]{\begingroup \@sanitize@url \@href}%
\providecommand \@href[1]{\@@startlink{#1}\@@href}%
\providecommand \@@href[1]{\endgroup#1\@@endlink}%
\providecommand \@sanitize@url [0]{\catcode `\\12\catcode `\$12\catcode
  `\&12\catcode `\#12\catcode `\^12\catcode `\_12\catcode `\%12\relax}%
\providecommand \@@startlink[1]{}%
\providecommand \@@endlink[0]{}%
\providecommand \url  [0]{\begingroup\@sanitize@url \@url }%
\providecommand \@url [1]{\endgroup\@href {#1}{\urlprefix }}%
\providecommand \urlprefix  [0]{URL }%
\providecommand \Eprint [0]{\href }%
\providecommand \doibase [0]{http://dx.doi.org/}%
\providecommand \selectlanguage [0]{\@gobble}%
\providecommand \bibinfo  [0]{\@secondoftwo}%
\providecommand \bibfield  [0]{\@secondoftwo}%
\providecommand \translation [1]{[#1]}%
\providecommand \BibitemOpen [0]{}%
\providecommand \bibitemStop [0]{}%
\providecommand \bibitemNoStop [0]{.\EOS\space}%
\providecommand \EOS [0]{\spacefactor3000\relax}%
\providecommand \BibitemShut  [1]{\csname bibitem#1\endcsname}%
\let\auto@bib@innerbib\@empty
%</preamble>
\bibitem [{\citenamefont {Hawking}(1975)}]{Hawking:1974sw}%
  \BibitemOpen
  \bibfield  {author} {\bibinfo {author} {\bibfnamefont {S.~W.}\ \bibnamefont
  {Hawking}},\ }\href {\doibase 10.1007/BF02345020} {\bibfield  {journal}
  {\bibinfo  {journal} {Commun. Math. Phys.}\ }\textbf {\bibinfo {volume}
  {43}},\ \bibinfo {pages} {199} (\bibinfo {year} {1975})}\BibitemShut
  {NoStop}%
%%CITATION = CMPHA,43,199;%%
\bibitem [{\citenamefont {Hawking}(1976)}]{Hawking:1976ra}%
  \BibitemOpen
  \bibfield  {author} {\bibinfo {author} {\bibfnamefont {S.~W.}\ \bibnamefont
  {Hawking}},\ }\href {\doibase 10.1103/PhysRevD.14.2460} {\bibfield  {journal}
  {\bibinfo  {journal} {Phys. Rev.}\ }\textbf {\bibinfo {volume} {D14}},\
  \bibinfo {pages} {2460} (\bibinfo {year} {1976})}\BibitemShut {NoStop}%
%%CITATION = PHRVA,D14,2460;%%
\bibitem [{\citenamefont {Hawking}(1981)}]{Hawking:1980ng}%
  \BibitemOpen
  \bibfield  {author} {\bibinfo {author} {\bibfnamefont {S.~W.}\ \bibnamefont
  {Hawking}},\ }\href {\doibase 10.1007/BF01208279} {\bibfield  {journal}
  {\bibinfo  {journal} {Commun. Math. Phys.}\ }\textbf {\bibinfo {volume}
  {80}},\ \bibinfo {pages} {421} (\bibinfo {year} {1981})}\BibitemShut
  {NoStop}%
%%CITATION = CMPHA,80,421;%%
\bibitem [{\citenamefont {Hiscock}(1987)}]{Hiscock:1987hn}%
  \BibitemOpen
  \bibfield  {author} {\bibinfo {author} {\bibfnamefont {W.~A.}\ \bibnamefont
  {Hiscock}},\ }\href {\doibase 10.1103/PhysRevD.35.1161} {\bibfield  {journal}
  {\bibinfo  {journal} {Phys. Rev.}\ }\textbf {\bibinfo {volume} {D35}},\
  \bibinfo {pages} {1161} (\bibinfo {year} {1987})}\BibitemShut {NoStop}%
%%CITATION = PHRVA,D35,1161;%%
\bibitem [{\citenamefont {Berezin}, \citenamefont {Kuzmin},\ and\ \citenamefont
  {Tkachev}(1988)}]{Berezin:1987ea}%
  \BibitemOpen
  \bibfield  {author} {\bibinfo {author} {\bibfnamefont {V.~A.}\ \bibnamefont
  {Berezin}}, \bibinfo {author} {\bibfnamefont {V.~A.}\ \bibnamefont {Kuzmin}},
  \ and\ \bibinfo {author} {\bibfnamefont {I.~I.}\ \bibnamefont {Tkachev}},\
  }\href {\doibase 10.1016/0370-2693(88)90672-7} {\bibfield  {journal}
  {\bibinfo  {journal} {Phys. Lett.}\ }\textbf {\bibinfo {volume} {B207}},\
  \bibinfo {pages} {397} (\bibinfo {year} {1988})}\BibitemShut {NoStop}%
%%CITATION = PHLTA,B207,397;%%
\bibitem [{\citenamefont {Arnold}(1990)}]{Arnold:1989cq}%
  \BibitemOpen
  \bibfield  {author} {\bibinfo {author} {\bibfnamefont {P.~B.}\ \bibnamefont
  {Arnold}},\ }\href {\doibase 10.1016/0550-3213(90)90243-7} {\bibfield
  {journal} {\bibinfo  {journal} {Nucl. Phys.}\ }\textbf {\bibinfo {volume}
  {B346}},\ \bibinfo {pages} {160} (\bibinfo {year} {1990})}\BibitemShut
  {NoStop}%
%%CITATION = NUPHA,B346,160;%%
\bibitem [{\citenamefont {Aad}\ \emph {et~al.}(2015)\citenamefont {Aad} \emph
  {et~al.}}]{Aad:2015zhl}%
  \BibitemOpen
  \bibfield  {author} {\bibinfo {author} {\bibfnamefont {G.}~\bibnamefont
  {Aad}} \emph {et~al.} (\bibinfo {collaboration} {ATLAS, CMS}),\ }\bibfield
  {booktitle} {\emph {\bibinfo {booktitle} {{Proceedings, Meeting of the APS
  Division of Particles and Fields (DPF 2015): Ann Arbor, Michigan, USA, 4-8
  Aug 2015}}},\ }\href {\doibase 10.1103/PhysRevLett.114.191803} {\bibfield
  {journal} {\bibinfo  {journal} {Phys. Rev. Lett.}\ }\textbf {\bibinfo
  {volume} {114}},\ \bibinfo {pages} {191803} (\bibinfo {year} {2015})},\
  \Eprint {http://arxiv.org/abs/1503.07589} {arXiv:1503.07589 [hep-ex]}
  \BibitemShut {NoStop}%
%%CITATION = ARXIV:1503.07589;%%
\bibitem [{\citenamefont {Aad}\ \emph {et~al.}(2013)\citenamefont {Aad} \emph
  {et~al.}}]{Aad:2013wqa}%
  \BibitemOpen
  \bibfield  {author} {\bibinfo {author} {\bibfnamefont {G.}~\bibnamefont
  {Aad}} \emph {et~al.} (\bibinfo {collaboration} {ATLAS}),\ }\href {\doibase
  10.1016/j.physletb.2014.05.011, 10.1016/j.physletb.2013.08.010} {\bibfield
  {journal} {\bibinfo  {journal} {Phys. Lett.}\ }\textbf {\bibinfo {volume}
  {B726}},\ \bibinfo {pages} {88} (\bibinfo {year} {2013})},\ \bibinfo {note}
  {[Erratum: Phys. Lett.B734,406(2014)]},\ \Eprint
  {http://arxiv.org/abs/1307.1427} {arXiv:1307.1427 [hep-ex]} \BibitemShut
  {NoStop}%
%%CITATION = ARXIV:1307.1427;%%
\bibitem [{\citenamefont {Chatrchyan}\ \emph {et~al.}(2014)\citenamefont
  {Chatrchyan} \emph {et~al.}}]{Chatrchyan:2013mxa}%
  \BibitemOpen
  \bibfield  {author} {\bibinfo {author} {\bibfnamefont {S.}~\bibnamefont
  {Chatrchyan}} \emph {et~al.} (\bibinfo {collaboration} {CMS}),\ }\href
  {\doibase 10.1103/PhysRevD.89.092007} {\bibfield  {journal} {\bibinfo
  {journal} {Phys. Rev.}\ }\textbf {\bibinfo {volume} {D89}},\ \bibinfo {pages}
  {092007} (\bibinfo {year} {2014})},\ \Eprint {http://arxiv.org/abs/1312.5353}
  {arXiv:1312.5353 [hep-ex]} \BibitemShut {NoStop}%
%%CITATION = ARXIV:1312.5353;%%
\bibitem [{\citenamefont {Giardino}\ \emph {et~al.}(2014)\citenamefont
  {Giardino}, \citenamefont {Kannike}, \citenamefont {Masina}, \citenamefont
  {Raidal},\ and\ \citenamefont {Strumia}}]{Giardino:2013bma}%
  \BibitemOpen
  \bibfield  {author} {\bibinfo {author} {\bibfnamefont {P.~P.}\ \bibnamefont
  {Giardino}}, \bibinfo {author} {\bibfnamefont {K.}~\bibnamefont {Kannike}},
  \bibinfo {author} {\bibfnamefont {I.}~\bibnamefont {Masina}}, \bibinfo
  {author} {\bibfnamefont {M.}~\bibnamefont {Raidal}}, \ and\ \bibinfo {author}
  {\bibfnamefont {A.}~\bibnamefont {Strumia}},\ }\href {\doibase
  10.1007/JHEP05(2014)046} {\bibfield  {journal} {\bibinfo  {journal} {JHEP}\
  }\textbf {\bibinfo {volume} {05}},\ \bibinfo {pages} {046} (\bibinfo {year}
  {2014})},\ \Eprint {http://arxiv.org/abs/1303.3570} {arXiv:1303.3570
  [hep-ph]} \BibitemShut {NoStop}%
%%CITATION = ARXIV:1303.3570;%%
\bibitem [{\citenamefont {Khachatryan}\ \emph {et~al.}(2016)\citenamefont
  {Khachatryan} \emph {et~al.}}]{Khachatryan:2015hba}%
  \BibitemOpen
  \bibfield  {author} {\bibinfo {author} {\bibfnamefont {V.}~\bibnamefont
  {Khachatryan}} \emph {et~al.} (\bibinfo {collaboration} {CMS}),\ }\href
  {\doibase 10.1103/PhysRevD.93.072004} {\bibfield  {journal} {\bibinfo
  {journal} {Phys. Rev.}\ }\textbf {\bibinfo {volume} {D93}},\ \bibinfo {pages}
  {072004} (\bibinfo {year} {2016})},\ \Eprint
  {http://arxiv.org/abs/1509.04044} {arXiv:1509.04044 [hep-ex]} \BibitemShut
  {NoStop}%
%%CITATION = ARXIV:1509.04044;%%
\bibitem [{\citenamefont {Buttazzo}\ \emph {et~al.}(2013)\citenamefont
  {Buttazzo}, \citenamefont {Degrassi}, \citenamefont {Giardino}, \citenamefont
  {Giudice}, \citenamefont {Sala}, \citenamefont {Salvio},\ and\ \citenamefont
  {Strumia}}]{Buttazzo:2013uya}%
  \BibitemOpen
  \bibfield  {author} {\bibinfo {author} {\bibfnamefont {D.}~\bibnamefont
  {Buttazzo}}, \bibinfo {author} {\bibfnamefont {G.}~\bibnamefont {Degrassi}},
  \bibinfo {author} {\bibfnamefont {P.~P.}\ \bibnamefont {Giardino}}, \bibinfo
  {author} {\bibfnamefont {G.~F.}\ \bibnamefont {Giudice}}, \bibinfo {author}
  {\bibfnamefont {F.}~\bibnamefont {Sala}}, \bibinfo {author} {\bibfnamefont
  {A.}~\bibnamefont {Salvio}}, \ and\ \bibinfo {author} {\bibfnamefont
  {A.}~\bibnamefont {Strumia}},\ }\href {\doibase 10.1007/JHEP12(2013)089}
  {\bibfield  {journal} {\bibinfo  {journal} {JHEP}\ }\textbf {\bibinfo
  {volume} {12}},\ \bibinfo {pages} {089} (\bibinfo {year} {2013})},\ \Eprint
  {http://arxiv.org/abs/1307.3536} {arXiv:1307.3536 [hep-ph]} \BibitemShut
  {NoStop}%
%%CITATION = ARXIV:1307.3536;%%
\bibitem [{\citenamefont {Di~Luzio}\ and\ \citenamefont
  {Mihaila}(2014)}]{DiLuzio:2014bua}%
  \BibitemOpen
  \bibfield  {author} {\bibinfo {author} {\bibfnamefont {L.}~\bibnamefont
  {Di~Luzio}}\ and\ \bibinfo {author} {\bibfnamefont {L.}~\bibnamefont
  {Mihaila}},\ }\href {\doibase 10.1007/JHEP06(2014)079} {\bibfield  {journal}
  {\bibinfo  {journal} {JHEP}\ }\textbf {\bibinfo {volume} {06}},\ \bibinfo
  {pages} {079} (\bibinfo {year} {2014})},\ \Eprint
  {http://arxiv.org/abs/1404.7450} {arXiv:1404.7450 [hep-ph]} \BibitemShut
  {NoStop}%
%%CITATION = ARXIV:1404.7450;%%
\bibitem [{\citenamefont {Andreassen}, \citenamefont {Frost},\ and\
  \citenamefont {Schwartz}(2015)}]{Andreassen:2014eha}%
  \BibitemOpen
  \bibfield  {author} {\bibinfo {author} {\bibfnamefont {A.}~\bibnamefont
  {Andreassen}}, \bibinfo {author} {\bibfnamefont {W.}~\bibnamefont {Frost}}, \
  and\ \bibinfo {author} {\bibfnamefont {M.~D.}\ \bibnamefont {Schwartz}},\
  }\href {\doibase 10.1103/PhysRevD.91.016009} {\bibfield  {journal} {\bibinfo
  {journal} {Phys. Rev.}\ }\textbf {\bibinfo {volume} {D91}},\ \bibinfo {pages}
  {016009} (\bibinfo {year} {2015})},\ \Eprint {http://arxiv.org/abs/1408.0287}
  {arXiv:1408.0287 [hep-ph]} \BibitemShut {NoStop}%
%%CITATION = ARXIV:1408.0287;%%
\bibitem [{\citenamefont {Andreassen}, \citenamefont {Frost},\ and\
  \citenamefont {Schwartz}(2014)}]{Andreassen:2014gha}%
  \BibitemOpen
  \bibfield  {author} {\bibinfo {author} {\bibfnamefont {A.}~\bibnamefont
  {Andreassen}}, \bibinfo {author} {\bibfnamefont {W.}~\bibnamefont {Frost}}, \
  and\ \bibinfo {author} {\bibfnamefont {M.~D.}\ \bibnamefont {Schwartz}},\
  }\href {\doibase 10.1103/PhysRevLett.113.241801} {\bibfield  {journal}
  {\bibinfo  {journal} {Phys. Rev. Lett.}\ }\textbf {\bibinfo {volume} {113}},\
  \bibinfo {pages} {241801} (\bibinfo {year} {2014})},\ \Eprint
  {http://arxiv.org/abs/1408.0292} {arXiv:1408.0292 [hep-ph]} \BibitemShut
  {NoStop}%
%%CITATION = ARXIV:1408.0292;%%
\bibitem [{\citenamefont {Lalak}, \citenamefont {Lewicki},\ and\ \citenamefont
  {Olszewski}(2016)}]{Lalak:2016zlv}%
  \BibitemOpen
  \bibfield  {author} {\bibinfo {author} {\bibfnamefont {Z.}~\bibnamefont
  {Lalak}}, \bibinfo {author} {\bibfnamefont {M.}~\bibnamefont {Lewicki}}, \
  and\ \bibinfo {author} {\bibfnamefont {P.}~\bibnamefont {Olszewski}},\ }\href
  {\doibase 10.1103/PhysRevD.94.085028} {\bibfield  {journal} {\bibinfo
  {journal} {Phys. Rev.}\ }\textbf {\bibinfo {volume} {D94}},\ \bibinfo {pages}
  {085028} (\bibinfo {year} {2016})},\ \Eprint
  {http://arxiv.org/abs/1605.06713} {arXiv:1605.06713 [hep-ph]} \BibitemShut
  {NoStop}%
%%CITATION = ARXIV:1605.06713;%%
\bibitem [{\citenamefont {Espinosa}, \citenamefont {Garny},\ and\ \citenamefont
  {Konstandin}(2016)}]{Espinosa:2016uaw}%
  \BibitemOpen
  \bibfield  {author} {\bibinfo {author} {\bibfnamefont {J.~R.}\ \bibnamefont
  {Espinosa}}, \bibinfo {author} {\bibfnamefont {M.}~\bibnamefont {Garny}}, \
  and\ \bibinfo {author} {\bibfnamefont {T.}~\bibnamefont {Konstandin}},\
  }\href {\doibase 10.1103/PhysRevD.94.055026} {\bibfield  {journal} {\bibinfo
  {journal} {Phys. Rev.}\ }\textbf {\bibinfo {volume} {D94}},\ \bibinfo {pages}
  {055026} (\bibinfo {year} {2016})},\ \Eprint
  {http://arxiv.org/abs/1607.08432} {arXiv:1607.08432 [hep-ph]} \BibitemShut
  {NoStop}%
%%CITATION = ARXIV:1607.08432;%%
\bibitem [{\citenamefont {Espinosa}\ \emph {et~al.}(2017)\citenamefont
  {Espinosa}, \citenamefont {Garny}, \citenamefont {Konstandin},\ and\
  \citenamefont {Riotto}}]{Espinosa:2016nld}%
  \BibitemOpen
  \bibfield  {author} {\bibinfo {author} {\bibfnamefont {J.~R.}\ \bibnamefont
  {Espinosa}}, \bibinfo {author} {\bibfnamefont {M.}~\bibnamefont {Garny}},
  \bibinfo {author} {\bibfnamefont {T.}~\bibnamefont {Konstandin}}, \ and\
  \bibinfo {author} {\bibfnamefont {A.}~\bibnamefont {Riotto}},\ }\href
  {\doibase 10.1103/PhysRevD.95.056004} {\bibfield  {journal} {\bibinfo
  {journal} {Phys. Rev.}\ }\textbf {\bibinfo {volume} {D95}},\ \bibinfo {pages}
  {056004} (\bibinfo {year} {2017})},\ \Eprint
  {http://arxiv.org/abs/1608.06765} {arXiv:1608.06765 [hep-ph]} \BibitemShut
  {NoStop}%
%%CITATION = ARXIV:1608.06765;%%
\bibitem [{\citenamefont {Branchina}\ and\ \citenamefont
  {Messina}(2013)}]{Branchina:2013jra}%
  \BibitemOpen
  \bibfield  {author} {\bibinfo {author} {\bibfnamefont {V.}~\bibnamefont
  {Branchina}}\ and\ \bibinfo {author} {\bibfnamefont {E.}~\bibnamefont
  {Messina}},\ }\href {\doibase 10.1103/PhysRevLett.111.241801} {\bibfield
  {journal} {\bibinfo  {journal} {Phys. Rev. Lett.}\ }\textbf {\bibinfo
  {volume} {111}},\ \bibinfo {pages} {241801} (\bibinfo {year} {2013})},\
  \Eprint {http://arxiv.org/abs/1307.5193} {arXiv:1307.5193 [hep-ph]}
  \BibitemShut {NoStop}%
%%CITATION = ARXIV:1307.5193;%%
\bibitem [{\citenamefont {Lalak}, \citenamefont {Lewicki},\ and\ \citenamefont
  {Olszewski}(2014)}]{Lalak:2014qua}%
  \BibitemOpen
  \bibfield  {author} {\bibinfo {author} {\bibfnamefont {Z.}~\bibnamefont
  {Lalak}}, \bibinfo {author} {\bibfnamefont {M.}~\bibnamefont {Lewicki}}, \
  and\ \bibinfo {author} {\bibfnamefont {P.}~\bibnamefont {Olszewski}},\ }\href
  {\doibase 10.1007/JHEP05(2014)119} {\bibfield  {journal} {\bibinfo  {journal}
  {JHEP}\ }\textbf {\bibinfo {volume} {05}},\ \bibinfo {pages} {119} (\bibinfo
  {year} {2014})},\ \Eprint {http://arxiv.org/abs/1402.3826} {arXiv:1402.3826
  [hep-ph]} \BibitemShut {NoStop}%
%%CITATION = ARXIV:1402.3826;%%
\bibitem [{\citenamefont {Branchina}, \citenamefont {Messina},\ and\
  \citenamefont {Platania}(2014)}]{Branchina:2014usa}%
  \BibitemOpen
  \bibfield  {author} {\bibinfo {author} {\bibfnamefont {V.}~\bibnamefont
  {Branchina}}, \bibinfo {author} {\bibfnamefont {E.}~\bibnamefont {Messina}},
  \ and\ \bibinfo {author} {\bibfnamefont {A.}~\bibnamefont {Platania}},\
  }\href {\doibase 10.1007/JHEP09(2014)182} {\bibfield  {journal} {\bibinfo
  {journal} {JHEP}\ }\textbf {\bibinfo {volume} {09}},\ \bibinfo {pages} {182}
  (\bibinfo {year} {2014})},\ \Eprint {http://arxiv.org/abs/1407.4112}
  {arXiv:1407.4112 [hep-ph]} \BibitemShut {NoStop}%
%%CITATION = ARXIV:1407.4112;%%
\bibitem [{\citenamefont {Branchina}, \citenamefont {Messina},\ and\
  \citenamefont {Sher}(2015)}]{Branchina:2014rva}%
  \BibitemOpen
  \bibfield  {author} {\bibinfo {author} {\bibfnamefont {V.}~\bibnamefont
  {Branchina}}, \bibinfo {author} {\bibfnamefont {E.}~\bibnamefont {Messina}},
  \ and\ \bibinfo {author} {\bibfnamefont {M.}~\bibnamefont {Sher}},\ }\href
  {\doibase 10.1103/PhysRevD.91.013003} {\bibfield  {journal} {\bibinfo
  {journal} {Phys. Rev.}\ }\textbf {\bibinfo {volume} {D91}},\ \bibinfo {pages}
  {013003} (\bibinfo {year} {2015})},\ \Eprint {http://arxiv.org/abs/1408.5302}
  {arXiv:1408.5302 [hep-ph]} \BibitemShut {NoStop}%
%%CITATION = ARXIV:1408.5302;%%
\bibitem [{\citenamefont {Kobzarev}, \citenamefont {Okun},\ and\ \citenamefont
  {Voloshin}(1975)}]{Kobzarev:1974cp}%
  \BibitemOpen
  \bibfield  {author} {\bibinfo {author} {\bibfnamefont {I.~{\relax Yu}.}\
  \bibnamefont {Kobzarev}}, \bibinfo {author} {\bibfnamefont {L.~B.}\
  \bibnamefont {Okun}}, \ and\ \bibinfo {author} {\bibfnamefont {M.~B.}\
  \bibnamefont {Voloshin}},\ }\href@noop {} {\bibfield  {journal} {\bibinfo
  {journal} {Sov. J. Nucl. Phys.}\ }\textbf {\bibinfo {volume} {20}},\ \bibinfo
  {pages} {644} (\bibinfo {year} {1975})},\ \bibinfo {note} {[Yad.
  Fiz.20,1229(1974)]}\BibitemShut {NoStop}%
%%CITATION = SJNCA,20,644;%%
\bibitem [{\citenamefont {Coleman}(1977)}]{Coleman:1977py}%
  \BibitemOpen
  \bibfield  {author} {\bibinfo {author} {\bibfnamefont {S.~R.}\ \bibnamefont
  {Coleman}},\ }\href {\doibase 10.1103/PhysRevD.15.2929,
  10.1103/PhysRevD.16.1248} {\bibfield  {journal} {\bibinfo  {journal} {Phys.
  Rev.}\ }\textbf {\bibinfo {volume} {D15}},\ \bibinfo {pages} {2929} (\bibinfo
  {year} {1977})},\ \bibinfo {note} {[Erratum: Phys.
  Rev.D16,1248(1977)]}\BibitemShut {NoStop}%
%%CITATION = PHRVA,D15,2929;%%
\bibitem [{\citenamefont {Callan}\ and\ \citenamefont
  {Coleman}(1977)}]{Callan:1977pt}%
  \BibitemOpen
  \bibfield  {author} {\bibinfo {author} {\bibfnamefont {C.~G.}\ \bibnamefont
  {Callan}, \bibfnamefont {Jr.}}\ and\ \bibinfo {author} {\bibfnamefont
  {S.~R.}\ \bibnamefont {Coleman}},\ }\href {\doibase 10.1103/PhysRevD.16.1762}
  {\bibfield  {journal} {\bibinfo  {journal} {Phys. Rev.}\ }\textbf {\bibinfo
  {volume} {D16}},\ \bibinfo {pages} {1762} (\bibinfo {year}
  {1977})}\BibitemShut {NoStop}%
%%CITATION = PHRVA,D16,1762;%%
\bibitem [{\citenamefont {Degrassi}\ \emph {et~al.}(2012)\citenamefont
  {Degrassi}, \citenamefont {Di~Vita}, \citenamefont {Elias-Miro},
  \citenamefont {Espinosa}, \citenamefont {Giudice}, \citenamefont {Isidori},\
  and\ \citenamefont {Strumia}}]{Degrassi:2012ry}%
  \BibitemOpen
  \bibfield  {author} {\bibinfo {author} {\bibfnamefont {G.}~\bibnamefont
  {Degrassi}}, \bibinfo {author} {\bibfnamefont {S.}~\bibnamefont {Di~Vita}},
  \bibinfo {author} {\bibfnamefont {J.}~\bibnamefont {Elias-Miro}}, \bibinfo
  {author} {\bibfnamefont {J.~R.}\ \bibnamefont {Espinosa}}, \bibinfo {author}
  {\bibfnamefont {G.~F.}\ \bibnamefont {Giudice}}, \bibinfo {author}
  {\bibfnamefont {G.}~\bibnamefont {Isidori}}, \ and\ \bibinfo {author}
  {\bibfnamefont {A.}~\bibnamefont {Strumia}},\ }\href {\doibase
  10.1007/JHEP08(2012)098} {\bibfield  {journal} {\bibinfo  {journal} {JHEP}\
  }\textbf {\bibinfo {volume} {08}},\ \bibinfo {pages} {098} (\bibinfo {year}
  {2012})},\ \Eprint {http://arxiv.org/abs/1205.6497} {arXiv:1205.6497
  [hep-ph]} \BibitemShut {NoStop}%
%%CITATION = ARXIV:1205.6497;%%
\bibitem [{\citenamefont {Isidori}, \citenamefont {Ridolfi},\ and\
  \citenamefont {Strumia}(2001)}]{Isidori:2001bm}%
  \BibitemOpen
  \bibfield  {author} {\bibinfo {author} {\bibfnamefont {G.}~\bibnamefont
  {Isidori}}, \bibinfo {author} {\bibfnamefont {G.}~\bibnamefont {Ridolfi}}, \
  and\ \bibinfo {author} {\bibfnamefont {A.}~\bibnamefont {Strumia}},\ }\href
  {\doibase 10.1016/S0550-3213(01)00302-9} {\bibfield  {journal} {\bibinfo
  {journal} {Nucl. Phys.}\ }\textbf {\bibinfo {volume} {B609}},\ \bibinfo
  {pages} {387} (\bibinfo {year} {2001})},\ \Eprint
  {http://arxiv.org/abs/hep-ph/0104016} {arXiv:hep-ph/0104016 [hep-ph]}
  \BibitemShut {NoStop}%
%%CITATION = HEP-PH/0104016;%%
\bibitem [{\citenamefont {Ellis}\ \emph {et~al.}(2009)\citenamefont {Ellis},
  \citenamefont {Espinosa}, \citenamefont {Giudice}, \citenamefont {Hoecker},\
  and\ \citenamefont {Riotto}}]{Ellis:2009tp}%
  \BibitemOpen
  \bibfield  {author} {\bibinfo {author} {\bibfnamefont {J.}~\bibnamefont
  {Ellis}}, \bibinfo {author} {\bibfnamefont {J.~R.}\ \bibnamefont {Espinosa}},
  \bibinfo {author} {\bibfnamefont {G.~F.}\ \bibnamefont {Giudice}}, \bibinfo
  {author} {\bibfnamefont {A.}~\bibnamefont {Hoecker}}, \ and\ \bibinfo
  {author} {\bibfnamefont {A.}~\bibnamefont {Riotto}},\ }\href {\doibase
  10.1016/j.physletb.2009.07.054} {\bibfield  {journal} {\bibinfo  {journal}
  {Phys. Lett.}\ }\textbf {\bibinfo {volume} {B679}},\ \bibinfo {pages} {369}
  (\bibinfo {year} {2009})},\ \Eprint {http://arxiv.org/abs/0906.0954}
  {arXiv:0906.0954 [hep-ph]} \BibitemShut {NoStop}%
%%CITATION = ARXIV:0906.0954;%%
\bibitem [{\citenamefont {Elias-Miro}\ \emph {et~al.}(2012)\citenamefont
  {Elias-Miro}, \citenamefont {Espinosa}, \citenamefont {Giudice},
  \citenamefont {Isidori}, \citenamefont {Riotto},\ and\ \citenamefont
  {Strumia}}]{EliasMiro:2011aa}%
  \BibitemOpen
  \bibfield  {author} {\bibinfo {author} {\bibfnamefont {J.}~\bibnamefont
  {Elias-Miro}}, \bibinfo {author} {\bibfnamefont {J.~R.}\ \bibnamefont
  {Espinosa}}, \bibinfo {author} {\bibfnamefont {G.~F.}\ \bibnamefont
  {Giudice}}, \bibinfo {author} {\bibfnamefont {G.}~\bibnamefont {Isidori}},
  \bibinfo {author} {\bibfnamefont {A.}~\bibnamefont {Riotto}}, \ and\ \bibinfo
  {author} {\bibfnamefont {A.}~\bibnamefont {Strumia}},\ }\href {\doibase
  10.1016/j.physletb.2012.02.013} {\bibfield  {journal} {\bibinfo  {journal}
  {Phys. Lett.}\ }\textbf {\bibinfo {volume} {B709}},\ \bibinfo {pages} {222}
  (\bibinfo {year} {2012})},\ \Eprint {http://arxiv.org/abs/1112.3022}
  {arXiv:1112.3022 [hep-ph]} \BibitemShut {NoStop}%
%%CITATION = ARXIV:1112.3022;%%
\bibitem [{\citenamefont {Espinosa}, \citenamefont {Giudice},\ and\
  \citenamefont {Riotto}(2008)}]{Espinosa:2007qp}%
  \BibitemOpen
  \bibfield  {author} {\bibinfo {author} {\bibfnamefont {J.~R.}\ \bibnamefont
  {Espinosa}}, \bibinfo {author} {\bibfnamefont {G.~F.}\ \bibnamefont
  {Giudice}}, \ and\ \bibinfo {author} {\bibfnamefont {A.}~\bibnamefont
  {Riotto}},\ }\href {\doibase 10.1088/1475-7516/2008/05/002} {\bibfield
  {journal} {\bibinfo  {journal} {JCAP}\ }\textbf {\bibinfo {volume} {0805}},\
  \bibinfo {pages} {002} (\bibinfo {year} {2008})},\ \Eprint
  {http://arxiv.org/abs/0710.2484} {arXiv:0710.2484 [hep-ph]} \BibitemShut
  {NoStop}%
%%CITATION = ARXIV:0710.2484;%%
\bibitem [{\citenamefont {Herranen}\ \emph {et~al.}(2014)\citenamefont
  {Herranen}, \citenamefont {Markkanen}, \citenamefont {Nurmi},\ and\
  \citenamefont {Rajantie}}]{Herranen:2014cua}%
  \BibitemOpen
  \bibfield  {author} {\bibinfo {author} {\bibfnamefont {M.}~\bibnamefont
  {Herranen}}, \bibinfo {author} {\bibfnamefont {T.}~\bibnamefont {Markkanen}},
  \bibinfo {author} {\bibfnamefont {S.}~\bibnamefont {Nurmi}}, \ and\ \bibinfo
  {author} {\bibfnamefont {A.}~\bibnamefont {Rajantie}},\ }\href {\doibase
  10.1103/PhysRevLett.113.211102} {\bibfield  {journal} {\bibinfo  {journal}
  {Phys. Rev. Lett.}\ }\textbf {\bibinfo {volume} {113}},\ \bibinfo {pages}
  {211102} (\bibinfo {year} {2014})},\ \Eprint {http://arxiv.org/abs/1407.3141}
  {arXiv:1407.3141 [hep-ph]} \BibitemShut {NoStop}%
%%CITATION = ARXIV:1407.3141;%%
\bibitem [{\citenamefont {Hook}\ \emph {et~al.}(2015)\citenamefont {Hook},
  \citenamefont {Kearney}, \citenamefont {Shakya},\ and\ \citenamefont
  {Zurek}}]{Hook:2014uia}%
  \BibitemOpen
  \bibfield  {author} {\bibinfo {author} {\bibfnamefont {A.}~\bibnamefont
  {Hook}}, \bibinfo {author} {\bibfnamefont {J.}~\bibnamefont {Kearney}},
  \bibinfo {author} {\bibfnamefont {B.}~\bibnamefont {Shakya}}, \ and\ \bibinfo
  {author} {\bibfnamefont {K.~M.}\ \bibnamefont {Zurek}},\ }\href {\doibase
  10.1007/JHEP01(2015)061} {\bibfield  {journal} {\bibinfo  {journal} {JHEP}\
  }\textbf {\bibinfo {volume} {01}},\ \bibinfo {pages} {061} (\bibinfo {year}
  {2015})},\ \Eprint {http://arxiv.org/abs/1404.5953} {arXiv:1404.5953
  [hep-ph]} \BibitemShut {NoStop}%
%%CITATION = ARXIV:1404.5953;%%
\bibitem [{\citenamefont {Kearney}, \citenamefont {Yoo},\ and\ \citenamefont
  {Zurek}(2015)}]{Kearney:2015vba}%
  \BibitemOpen
  \bibfield  {author} {\bibinfo {author} {\bibfnamefont {J.}~\bibnamefont
  {Kearney}}, \bibinfo {author} {\bibfnamefont {H.}~\bibnamefont {Yoo}}, \ and\
  \bibinfo {author} {\bibfnamefont {K.~M.}\ \bibnamefont {Zurek}},\ }\href
  {\doibase 10.1103/PhysRevD.91.123537} {\bibfield  {journal} {\bibinfo
  {journal} {Phys. Rev.}\ }\textbf {\bibinfo {volume} {D91}},\ \bibinfo {pages}
  {123537} (\bibinfo {year} {2015})},\ \Eprint
  {http://arxiv.org/abs/1503.05193} {arXiv:1503.05193 [hep-th]} \BibitemShut
  {NoStop}%
%%CITATION = ARXIV:1503.05193;%%
\bibitem [{\citenamefont {Kohri}\ and\ \citenamefont
  {Matsui}(2016{\natexlab{a}})}]{Kohri:2016qqv}%
  \BibitemOpen
  \bibfield  {author} {\bibinfo {author} {\bibfnamefont {K.}~\bibnamefont
  {Kohri}}\ and\ \bibinfo {author} {\bibfnamefont {H.}~\bibnamefont {Matsui}},\
  }\href@noop {} {\  (\bibinfo {year} {2016}{\natexlab{a}})},\ \Eprint
  {http://arxiv.org/abs/1607.08133} {arXiv:1607.08133 [hep-ph]} \BibitemShut
  {NoStop}%
%%CITATION = ARXIV:1607.08133;%%
\bibitem [{\citenamefont {Kohri}\ and\ \citenamefont
  {Matsui}(2017)}]{Kohri:2017iyl}%
  \BibitemOpen
  \bibfield  {author} {\bibinfo {author} {\bibfnamefont {K.}~\bibnamefont
  {Kohri}}\ and\ \bibinfo {author} {\bibfnamefont {H.}~\bibnamefont {Matsui}},\
  }\href@noop {} {\  (\bibinfo {year} {2017})},\ \Eprint
  {http://arxiv.org/abs/1704.06884} {arXiv:1704.06884 [hep-ph]} \BibitemShut
  {NoStop}%
%%CITATION = ARXIV:1704.06884;%%
\bibitem [{\citenamefont {Czerwinska}\ \emph {et~al.}(2016)\citenamefont
  {Czerwinska}, \citenamefont {Lalak}, \citenamefont {Lewicki},\ and\
  \citenamefont {Olszewski}}]{Czerwinska:2016fky}%
  \BibitemOpen
  \bibfield  {author} {\bibinfo {author} {\bibfnamefont {O.}~\bibnamefont
  {Czerwinska}}, \bibinfo {author} {\bibfnamefont {Z.}~\bibnamefont {Lalak}},
  \bibinfo {author} {\bibfnamefont {M.}~\bibnamefont {Lewicki}}, \ and\
  \bibinfo {author} {\bibfnamefont {P.}~\bibnamefont {Olszewski}},\ }\href
  {\doibase 10.1007/JHEP10(2016)004} {\bibfield  {journal} {\bibinfo  {journal}
  {JHEP}\ }\textbf {\bibinfo {volume} {10}},\ \bibinfo {pages} {004} (\bibinfo
  {year} {2016})},\ \Eprint {http://arxiv.org/abs/1606.07808} {arXiv:1606.07808
  [hep-ph]} \BibitemShut {NoStop}%
%%CITATION = ARXIV:1606.07808;%%
\bibitem [{\citenamefont {Gong}\ and\ \citenamefont
  {Kitajima}(2017)}]{Gong:2017mwt}%
  \BibitemOpen
  \bibfield  {author} {\bibinfo {author} {\bibfnamefont {J.-O.}\ \bibnamefont
  {Gong}}\ and\ \bibinfo {author} {\bibfnamefont {N.}~\bibnamefont
  {Kitajima}},\ }\href@noop {} {\  (\bibinfo {year} {2017})},\ \Eprint
  {http://arxiv.org/abs/1705.11178} {arXiv:1705.11178 [hep-ph]} \BibitemShut
  {NoStop}%
%%CITATION = ARXIV:1705.11178;%%
\bibitem [{\citenamefont {East}\ \emph {et~al.}(2017)\citenamefont {East},
  \citenamefont {Kearney}, \citenamefont {Shakya}, \citenamefont {Yoo},\ and\
  \citenamefont {Zurek}}]{East:2016anr}%
  \BibitemOpen
  \bibfield  {author} {\bibinfo {author} {\bibfnamefont {W.~E.}\ \bibnamefont
  {East}}, \bibinfo {author} {\bibfnamefont {J.}~\bibnamefont {Kearney}},
  \bibinfo {author} {\bibfnamefont {B.}~\bibnamefont {Shakya}}, \bibinfo
  {author} {\bibfnamefont {H.}~\bibnamefont {Yoo}}, \ and\ \bibinfo {author}
  {\bibfnamefont {K.~M.}\ \bibnamefont {Zurek}},\ }\href {\doibase
  10.1103/PhysRevD.95.023526} {\bibfield  {journal} {\bibinfo  {journal} {Phys.
  Rev.}\ }\textbf {\bibinfo {volume} {D95}},\ \bibinfo {pages} {023526}
  (\bibinfo {year} {2017})},\ \bibinfo {note} {[Phys. Rev.D95,023526(2017)]},\
  \Eprint {http://arxiv.org/abs/1607.00381} {arXiv:1607.00381 [hep-ph]}
  \BibitemShut {NoStop}%
%%CITATION = ARXIV:1607.00381;%%
\bibitem [{\citenamefont {Espinosa}\ \emph {et~al.}(2015)\citenamefont
  {Espinosa}, \citenamefont {Giudice}, \citenamefont {Morgante}, \citenamefont
  {Riotto}, \citenamefont {Senatore}, \citenamefont {Strumia},\ and\
  \citenamefont {Tetradis}}]{Espinosa:2015qea}%
  \BibitemOpen
  \bibfield  {author} {\bibinfo {author} {\bibfnamefont {J.~R.}\ \bibnamefont
  {Espinosa}}, \bibinfo {author} {\bibfnamefont {G.~F.}\ \bibnamefont
  {Giudice}}, \bibinfo {author} {\bibfnamefont {E.}~\bibnamefont {Morgante}},
  \bibinfo {author} {\bibfnamefont {A.}~\bibnamefont {Riotto}}, \bibinfo
  {author} {\bibfnamefont {L.}~\bibnamefont {Senatore}}, \bibinfo {author}
  {\bibfnamefont {A.}~\bibnamefont {Strumia}}, \ and\ \bibinfo {author}
  {\bibfnamefont {N.}~\bibnamefont {Tetradis}},\ }\href {\doibase
  10.1007/JHEP09(2015)174} {\bibfield  {journal} {\bibinfo  {journal} {JHEP}\
  }\textbf {\bibinfo {volume} {09}},\ \bibinfo {pages} {174} (\bibinfo {year}
  {2015})},\ \Eprint {http://arxiv.org/abs/1505.04825} {arXiv:1505.04825
  [hep-ph]} \BibitemShut {NoStop}%
%%CITATION = ARXIV:1505.04825;%%
\bibitem [{\citenamefont {Joti}\ \emph {et~al.}(2017)\citenamefont {Joti},
  \citenamefont {Katsis}, \citenamefont {Loupas}, \citenamefont {Salvio},
  \citenamefont {Strumia}, \citenamefont {Tetradis},\ and\ \citenamefont
  {Urbano}}]{Joti:2017fwe}%
  \BibitemOpen
  \bibfield  {author} {\bibinfo {author} {\bibfnamefont {A.}~\bibnamefont
  {Joti}}, \bibinfo {author} {\bibfnamefont {A.}~\bibnamefont {Katsis}},
  \bibinfo {author} {\bibfnamefont {D.}~\bibnamefont {Loupas}}, \bibinfo
  {author} {\bibfnamefont {A.}~\bibnamefont {Salvio}}, \bibinfo {author}
  {\bibfnamefont {A.}~\bibnamefont {Strumia}}, \bibinfo {author} {\bibfnamefont
  {N.}~\bibnamefont {Tetradis}}, \ and\ \bibinfo {author} {\bibfnamefont
  {A.}~\bibnamefont {Urbano}},\ }\href@noop {} {\  (\bibinfo {year} {2017})},\
  \Eprint {http://arxiv.org/abs/1706.00792} {arXiv:1706.00792 [hep-ph]}
  \BibitemShut {NoStop}%
%%CITATION = ARXIV:1706.00792;%%
\bibitem [{\citenamefont {Herranen}\ \emph {et~al.}(2015)\citenamefont
  {Herranen}, \citenamefont {Markkanen}, \citenamefont {Nurmi},\ and\
  \citenamefont {Rajantie}}]{Herranen:2015ima}%
  \BibitemOpen
  \bibfield  {author} {\bibinfo {author} {\bibfnamefont {M.}~\bibnamefont
  {Herranen}}, \bibinfo {author} {\bibfnamefont {T.}~\bibnamefont {Markkanen}},
  \bibinfo {author} {\bibfnamefont {S.}~\bibnamefont {Nurmi}}, \ and\ \bibinfo
  {author} {\bibfnamefont {A.}~\bibnamefont {Rajantie}},\ }\href {\doibase
  10.1103/PhysRevLett.115.241301} {\bibfield  {journal} {\bibinfo  {journal}
  {Phys. Rev. Lett.}\ }\textbf {\bibinfo {volume} {115}},\ \bibinfo {pages}
  {241301} (\bibinfo {year} {2015})},\ \Eprint
  {http://arxiv.org/abs/1506.04065} {arXiv:1506.04065 [hep-ph]} \BibitemShut
  {NoStop}%
%%CITATION = ARXIV:1506.04065;%%
\bibitem [{\citenamefont {Kohri}\ and\ \citenamefont
  {Matsui}(2016{\natexlab{b}})}]{Kohri:2016wof}%
  \BibitemOpen
  \bibfield  {author} {\bibinfo {author} {\bibfnamefont {K.}~\bibnamefont
  {Kohri}}\ and\ \bibinfo {author} {\bibfnamefont {H.}~\bibnamefont {Matsui}},\
  }\href {\doibase 10.1103/PhysRevD.94.103509} {\bibfield  {journal} {\bibinfo
  {journal} {Phys. Rev.}\ }\textbf {\bibinfo {volume} {D94}},\ \bibinfo {pages}
  {103509} (\bibinfo {year} {2016}{\natexlab{b}})},\ \Eprint
  {http://arxiv.org/abs/1602.02100} {arXiv:1602.02100 [hep-ph]} \BibitemShut
  {NoStop}%
%%CITATION = ARXIV:1602.02100;%%
\bibitem [{\citenamefont {Ema}, \citenamefont {Mukaida},\ and\ \citenamefont
  {Nakayama}(2016)}]{Ema:2016kpf}%
  \BibitemOpen
  \bibfield  {author} {\bibinfo {author} {\bibfnamefont {Y.}~\bibnamefont
  {Ema}}, \bibinfo {author} {\bibfnamefont {K.}~\bibnamefont {Mukaida}}, \ and\
  \bibinfo {author} {\bibfnamefont {K.}~\bibnamefont {Nakayama}},\ }\href
  {\doibase 10.1088/1475-7516/2016/10/043} {\bibfield  {journal} {\bibinfo
  {journal} {JCAP}\ }\textbf {\bibinfo {volume} {1610}},\ \bibinfo {pages}
  {043} (\bibinfo {year} {2016})},\ \Eprint {http://arxiv.org/abs/1602.00483}
  {arXiv:1602.00483 [hep-ph]} \BibitemShut {NoStop}%
%%CITATION = ARXIV:1602.00483;%%
\bibitem [{\citenamefont {Enqvist}\ \emph {et~al.}(2016)\citenamefont
  {Enqvist}, \citenamefont {Karciauskas}, \citenamefont {Lebedev},
  \citenamefont {Rusak},\ and\ \citenamefont {Zatta}}]{Enqvist:2016mqj}%
  \BibitemOpen
  \bibfield  {author} {\bibinfo {author} {\bibfnamefont {K.}~\bibnamefont
  {Enqvist}}, \bibinfo {author} {\bibfnamefont {M.}~\bibnamefont
  {Karciauskas}}, \bibinfo {author} {\bibfnamefont {O.}~\bibnamefont
  {Lebedev}}, \bibinfo {author} {\bibfnamefont {S.}~\bibnamefont {Rusak}}, \
  and\ \bibinfo {author} {\bibfnamefont {M.}~\bibnamefont {Zatta}},\ }\href
  {\doibase 10.1088/1475-7516/2016/11/025} {\bibfield  {journal} {\bibinfo
  {journal} {JCAP}\ }\textbf {\bibinfo {volume} {1611}},\ \bibinfo {pages}
  {025} (\bibinfo {year} {2016})},\ \Eprint {http://arxiv.org/abs/1608.08848}
  {arXiv:1608.08848 [hep-ph]} \BibitemShut {NoStop}%
%%CITATION = ARXIV:1608.08848;%%
\bibitem [{\citenamefont {Postma}\ and\ \citenamefont {van~de
  Vis}(2017)}]{Postma:2017hbk}%
  \BibitemOpen
  \bibfield  {author} {\bibinfo {author} {\bibfnamefont {M.}~\bibnamefont
  {Postma}}\ and\ \bibinfo {author} {\bibfnamefont {J.}~\bibnamefont {van~de
  Vis}},\ }\href {\doibase 10.1088/1475-7516/2017/05/004} {\bibfield  {journal}
  {\bibinfo  {journal} {JCAP}\ }\textbf {\bibinfo {volume} {1705}},\ \bibinfo
  {pages} {004} (\bibinfo {year} {2017})},\ \Eprint
  {http://arxiv.org/abs/1702.07636} {arXiv:1702.07636 [hep-ph]} \BibitemShut
  {NoStop}%
%%CITATION = ARXIV:1702.07636;%%
\bibitem [{\citenamefont {Ema}\ \emph {et~al.}(2017)\citenamefont {Ema},
  \citenamefont {Karciauskas}, \citenamefont {Lebedev},\ and\ \citenamefont
  {Zatta}}]{Ema:2017loe}%
  \BibitemOpen
  \bibfield  {author} {\bibinfo {author} {\bibfnamefont {Y.}~\bibnamefont
  {Ema}}, \bibinfo {author} {\bibfnamefont {M.}~\bibnamefont {Karciauskas}},
  \bibinfo {author} {\bibfnamefont {O.}~\bibnamefont {Lebedev}}, \ and\
  \bibinfo {author} {\bibfnamefont {M.}~\bibnamefont {Zatta}},\ }\href@noop {}
  {\  (\bibinfo {year} {2017})},\ \Eprint {http://arxiv.org/abs/1703.04681}
  {arXiv:1703.04681 [hep-ph]} \BibitemShut {NoStop}%
%%CITATION = ARXIV:1703.04681;%%
\bibitem [{\citenamefont {Burda}, \citenamefont {Gregory},\ and\ \citenamefont
  {Moss}(2015{\natexlab{a}})}]{Burda:2015isa}%
  \BibitemOpen
  \bibfield  {author} {\bibinfo {author} {\bibfnamefont {P.}~\bibnamefont
  {Burda}}, \bibinfo {author} {\bibfnamefont {R.}~\bibnamefont {Gregory}}, \
  and\ \bibinfo {author} {\bibfnamefont {I.}~\bibnamefont {Moss}},\ }\href
  {\doibase 10.1103/PhysRevLett.115.071303} {\bibfield  {journal} {\bibinfo
  {journal} {Phys. Rev. Lett.}\ }\textbf {\bibinfo {volume} {115}},\ \bibinfo
  {pages} {071303} (\bibinfo {year} {2015}{\natexlab{a}})},\ \Eprint
  {http://arxiv.org/abs/1501.04937} {arXiv:1501.04937 [hep-th]} \BibitemShut
  {NoStop}%
%%CITATION = ARXIV:1501.04937;%%
\bibitem [{\citenamefont {Burda}, \citenamefont {Gregory},\ and\ \citenamefont
  {Moss}(2015{\natexlab{b}})}]{Burda:2015yfa}%
  \BibitemOpen
  \bibfield  {author} {\bibinfo {author} {\bibfnamefont {P.}~\bibnamefont
  {Burda}}, \bibinfo {author} {\bibfnamefont {R.}~\bibnamefont {Gregory}}, \
  and\ \bibinfo {author} {\bibfnamefont {I.}~\bibnamefont {Moss}},\ }\href
  {\doibase 10.1007/JHEP08(2015)114} {\bibfield  {journal} {\bibinfo  {journal}
  {JHEP}\ }\textbf {\bibinfo {volume} {08}},\ \bibinfo {pages} {114} (\bibinfo
  {year} {2015}{\natexlab{b}})},\ \Eprint {http://arxiv.org/abs/1503.07331}
  {arXiv:1503.07331 [hep-th]} \BibitemShut {NoStop}%
%%CITATION = ARXIV:1503.07331;%%
\bibitem [{\citenamefont {Burda}, \citenamefont {Gregory},\ and\ \citenamefont
  {Moss}(2016)}]{Burda:2016mou}%
  \BibitemOpen
  \bibfield  {author} {\bibinfo {author} {\bibfnamefont {P.}~\bibnamefont
  {Burda}}, \bibinfo {author} {\bibfnamefont {R.}~\bibnamefont {Gregory}}, \
  and\ \bibinfo {author} {\bibfnamefont {I.}~\bibnamefont {Moss}},\ }\href
  {\doibase 10.1007/JHEP06(2016)025} {\bibfield  {journal} {\bibinfo  {journal}
  {JHEP}\ }\textbf {\bibinfo {volume} {06}},\ \bibinfo {pages} {025} (\bibinfo
  {year} {2016})},\ \Eprint {http://arxiv.org/abs/1601.02152} {arXiv:1601.02152
  [hep-th]} \BibitemShut {NoStop}%
%%CITATION = ARXIV:1601.02152;%%
\bibitem [{\citenamefont {Grinstein}\ and\ \citenamefont
  {Murphy}(2015)}]{Grinstein:2015jda}%
  \BibitemOpen
  \bibfield  {author} {\bibinfo {author} {\bibfnamefont {B.}~\bibnamefont
  {Grinstein}}\ and\ \bibinfo {author} {\bibfnamefont {C.~W.}\ \bibnamefont
  {Murphy}},\ }\href {\doibase 10.1007/JHEP12(2015)063} {\bibfield  {journal}
  {\bibinfo  {journal} {JHEP}\ }\textbf {\bibinfo {volume} {12}},\ \bibinfo
  {pages} {063} (\bibinfo {year} {2015})},\ \Eprint
  {http://arxiv.org/abs/1509.05405} {arXiv:1509.05405 [hep-ph]} \BibitemShut
  {NoStop}%
%%CITATION = ARXIV:1509.05405;%%
\bibitem [{\citenamefont {Tetradis}(2016)}]{Tetradis:2016vqb}%
  \BibitemOpen
  \bibfield  {author} {\bibinfo {author} {\bibfnamefont {N.}~\bibnamefont
  {Tetradis}},\ }\href {\doibase 10.1088/1475-7516/2016/09/036} {\bibfield
  {journal} {\bibinfo  {journal} {JCAP}\ }\textbf {\bibinfo {volume} {1609}},\
  \bibinfo {pages} {036} (\bibinfo {year} {2016})},\ \Eprint
  {http://arxiv.org/abs/1606.04018} {arXiv:1606.04018 [hep-ph]} \BibitemShut
  {NoStop}%
%%CITATION = ARXIV:1606.04018;%%
\bibitem [{\citenamefont {Cheung}\ and\ \citenamefont
  {Leichenauer}(2014)}]{Cheung:2013sxa}%
  \BibitemOpen
  \bibfield  {author} {\bibinfo {author} {\bibfnamefont {C.}~\bibnamefont
  {Cheung}}\ and\ \bibinfo {author} {\bibfnamefont {S.}~\bibnamefont
  {Leichenauer}},\ }\href {\doibase 10.1103/PhysRevD.89.104035} {\bibfield
  {journal} {\bibinfo  {journal} {Phys. Rev.}\ }\textbf {\bibinfo {volume}
  {D89}},\ \bibinfo {pages} {104035} (\bibinfo {year} {2014})},\ \Eprint
  {http://arxiv.org/abs/1309.0530} {arXiv:1309.0530 [hep-ph]} \BibitemShut
  {NoStop}%
%%CITATION = ARXIV:1309.0530;%%
\bibitem [{\citenamefont {Canko}\ \emph {et~al.}(2017)\citenamefont {Canko},
  \citenamefont {Gialamas}, \citenamefont {Jelic-Cizmek}, \citenamefont
  {Riotto},\ and\ \citenamefont {Tetradis}}]{Canko:2017ebb}%
  \BibitemOpen
  \bibfield  {author} {\bibinfo {author} {\bibfnamefont {D.}~\bibnamefont
  {Canko}}, \bibinfo {author} {\bibfnamefont {I.}~\bibnamefont {Gialamas}},
  \bibinfo {author} {\bibfnamefont {G.}~\bibnamefont {Jelic-Cizmek}}, \bibinfo
  {author} {\bibfnamefont {A.}~\bibnamefont {Riotto}}, \ and\ \bibinfo {author}
  {\bibfnamefont {N.}~\bibnamefont {Tetradis}},\ }\href@noop {} {\  (\bibinfo
  {year} {2017})},\ \Eprint {http://arxiv.org/abs/1706.01364} {arXiv:1706.01364
  [hep-th]} \BibitemShut {NoStop}%
%%CITATION = ARXIV:1706.01364;%%
\bibitem [{\citenamefont {Gorbunov}, \citenamefont {Levkov},\ and\
  \citenamefont {Panin}(2017)}]{Gorbunov:2017fhq}%
  \BibitemOpen
  \bibfield  {author} {\bibinfo {author} {\bibfnamefont {D.}~\bibnamefont
  {Gorbunov}}, \bibinfo {author} {\bibfnamefont {D.}~\bibnamefont {Levkov}}, \
  and\ \bibinfo {author} {\bibfnamefont {A.}~\bibnamefont {Panin}},\
  }\href@noop {} {\  (\bibinfo {year} {2017})},\ \Eprint
  {http://arxiv.org/abs/1704.05399} {arXiv:1704.05399 [astro-ph.CO]}
  \BibitemShut {NoStop}%
%%CITATION = ARXIV:1704.05399;%%
\bibitem [{\citenamefont {Mukaida}\ and\ \citenamefont
  {Yamada}(2017)}]{Mukaida:2017bgd}%
  \BibitemOpen
  \bibfield  {author} {\bibinfo {author} {\bibfnamefont {K.}~\bibnamefont
  {Mukaida}}\ and\ \bibinfo {author} {\bibfnamefont {M.}~\bibnamefont
  {Yamada}},\ }\href@noop {} {\  (\bibinfo {year} {2017})},\ \Eprint
  {http://arxiv.org/abs/1706.04523} {arXiv:1706.04523 [hep-th]} \BibitemShut
  {NoStop}%
%%CITATION = ARXIV:1706.04523;%%
\bibitem [{\citenamefont {Hawking}(1971)}]{Hawking:1971ei}%
  \BibitemOpen
  \bibfield  {author} {\bibinfo {author} {\bibfnamefont {S.}~\bibnamefont
  {Hawking}},\ }\href@noop {} {\bibfield  {journal} {\bibinfo  {journal} {Mon.
  Not. Roy. Astron. Soc.}\ }\textbf {\bibinfo {volume} {152}},\ \bibinfo
  {pages} {75} (\bibinfo {year} {1971})}\BibitemShut {NoStop}%
%%CITATION = MNRAA,152,75;%%
\bibitem [{\citenamefont {Carr}\ and\ \citenamefont
  {Hawking}(1974)}]{Carr:1974nx}%
  \BibitemOpen
  \bibfield  {author} {\bibinfo {author} {\bibfnamefont {B.~J.}\ \bibnamefont
  {Carr}}\ and\ \bibinfo {author} {\bibfnamefont {S.~W.}\ \bibnamefont
  {Hawking}},\ }\href@noop {} {\bibfield  {journal} {\bibinfo  {journal} {Mon.
  Not. Roy. Astron. Soc.}\ }\textbf {\bibinfo {volume} {168}},\ \bibinfo
  {pages} {399} (\bibinfo {year} {1974})}\BibitemShut {NoStop}%
%%CITATION = MNRAA,168,399;%%
\bibitem [{\citenamefont {Carr}(1975)}]{Carr:1975qj}%
  \BibitemOpen
  \bibfield  {author} {\bibinfo {author} {\bibfnamefont {B.~J.}\ \bibnamefont
  {Carr}},\ }\href {\doibase 10.1086/153853} {\bibfield  {journal} {\bibinfo
  {journal} {Astrophys. J.}\ }\textbf {\bibinfo {volume} {201}},\ \bibinfo
  {pages} {1} (\bibinfo {year} {1975})}\BibitemShut {NoStop}%
%%CITATION = ASJOA,201,1;%%
\bibitem [{\citenamefont {Cole}\ and\ \citenamefont
  {Byrnes}(2017)}]{Cole:2017gle}%
  \BibitemOpen
  \bibfield  {author} {\bibinfo {author} {\bibfnamefont {P.~S.}\ \bibnamefont
  {Cole}}\ and\ \bibinfo {author} {\bibfnamefont {C.~T.}\ \bibnamefont
  {Byrnes}},\ }\href@noop {} {\  (\bibinfo {year} {2017})},\ \Eprint
  {http://arxiv.org/abs/1706.10288} {arXiv:1706.10288 [astro-ph.CO]}
  \BibitemShut {NoStop}%
%%CITATION = ARXIV:1706.10288;%%
\bibitem [{\citenamefont {Coleman}\ and\ \citenamefont
  {De~Luccia}(1980)}]{Coleman:1980aw}%
  \BibitemOpen
  \bibfield  {author} {\bibinfo {author} {\bibfnamefont {S.~R.}\ \bibnamefont
  {Coleman}}\ and\ \bibinfo {author} {\bibfnamefont {F.}~\bibnamefont
  {De~Luccia}},\ }\href {\doibase 10.1103/PhysRevD.21.3305} {\bibfield
  {journal} {\bibinfo  {journal} {Phys. Rev.}\ }\textbf {\bibinfo {volume}
  {D21}},\ \bibinfo {pages} {3305} (\bibinfo {year} {1980})}\BibitemShut
  {NoStop}%
%%CITATION = PHRVA,D21,3305;%%
\bibitem [{\citenamefont {Hawking}\ and\ \citenamefont
  {Moss}(1982)}]{Hawking:1981fz}%
  \BibitemOpen
  \bibfield  {author} {\bibinfo {author} {\bibfnamefont {S.~W.}\ \bibnamefont
  {Hawking}}\ and\ \bibinfo {author} {\bibfnamefont {I.~G.}\ \bibnamefont
  {Moss}},\ }\href {\doibase 10.1016/0370-2693(82)90946-7} {\bibfield
  {journal} {\bibinfo  {journal} {Phys. Lett.}\ }\textbf {\bibinfo {volume}
  {110B}},\ \bibinfo {pages} {35} (\bibinfo {year} {1982})}\BibitemShut
  {NoStop}%
%%CITATION = PHLTA,110B,35;%%
\bibitem [{\citenamefont {Arnold}\ and\ \citenamefont
  {Vokos}(1991)}]{Arnold:1991cv}%
  \BibitemOpen
  \bibfield  {author} {\bibinfo {author} {\bibfnamefont {P.~B.}\ \bibnamefont
  {Arnold}}\ and\ \bibinfo {author} {\bibfnamefont {S.}~\bibnamefont {Vokos}},\
  }\href {\doibase 10.1103/PhysRevD.44.3620} {\bibfield  {journal} {\bibinfo
  {journal} {Phys. Rev.}\ }\textbf {\bibinfo {volume} {D44}},\ \bibinfo {pages}
  {3620} (\bibinfo {year} {1991})}\BibitemShut {NoStop}%
%%CITATION = PHRVA,D44,3620;%%
\bibitem [{\citenamefont {Carrington}(1992)}]{Carrington:1991hz}%
  \BibitemOpen
  \bibfield  {author} {\bibinfo {author} {\bibfnamefont {M.~E.}\ \bibnamefont
  {Carrington}},\ }\href {\doibase 10.1103/PhysRevD.45.2933} {\bibfield
  {journal} {\bibinfo  {journal} {Phys. Rev.}\ }\textbf {\bibinfo {volume}
  {D45}},\ \bibinfo {pages} {2933} (\bibinfo {year} {1992})}\BibitemShut
  {NoStop}%
%%CITATION = PHRVA,D45,2933;%%
\bibitem [{\citenamefont {Anderson}\ and\ \citenamefont
  {Hall}(1992)}]{Anderson:1991zb}%
  \BibitemOpen
  \bibfield  {author} {\bibinfo {author} {\bibfnamefont {G.~W.}\ \bibnamefont
  {Anderson}}\ and\ \bibinfo {author} {\bibfnamefont {L.~J.}\ \bibnamefont
  {Hall}},\ }\href {\doibase 10.1103/PhysRevD.45.2685} {\bibfield  {journal}
  {\bibinfo  {journal} {Phys. Rev.}\ }\textbf {\bibinfo {volume} {D45}},\
  \bibinfo {pages} {2685} (\bibinfo {year} {1992})}\BibitemShut {NoStop}%
%%CITATION = PHRVA,D45,2685;%%
\bibitem [{\citenamefont {Delaunay}, \citenamefont {Grojean},\ and\
  \citenamefont {Wells}(2008)}]{Delaunay:2007wb}%
  \BibitemOpen
  \bibfield  {author} {\bibinfo {author} {\bibfnamefont {C.}~\bibnamefont
  {Delaunay}}, \bibinfo {author} {\bibfnamefont {C.}~\bibnamefont {Grojean}}, \
  and\ \bibinfo {author} {\bibfnamefont {J.~D.}\ \bibnamefont {Wells}},\ }\href
  {\doibase 10.1088/1126-6708/2008/04/029} {\bibfield  {journal} {\bibinfo
  {journal} {JHEP}\ }\textbf {\bibinfo {volume} {04}},\ \bibinfo {pages} {029}
  (\bibinfo {year} {2008})},\ \Eprint {http://arxiv.org/abs/0711.2511}
  {arXiv:0711.2511 [hep-ph]} \BibitemShut {NoStop}%
%%CITATION = ARXIV:0711.2511;%%
\bibitem [{\citenamefont {Delle~Rose}, \citenamefont {Marzo},\ and\
  \citenamefont {Urbano}(2016)}]{Rose:2015lna}%
  \BibitemOpen
  \bibfield  {author} {\bibinfo {author} {\bibfnamefont {L.}~\bibnamefont
  {Delle~Rose}}, \bibinfo {author} {\bibfnamefont {C.}~\bibnamefont {Marzo}}, \
  and\ \bibinfo {author} {\bibfnamefont {A.}~\bibnamefont {Urbano}},\ }\href
  {\doibase 10.1007/JHEP05(2016)050} {\bibfield  {journal} {\bibinfo  {journal}
  {JHEP}\ }\textbf {\bibinfo {volume} {05}},\ \bibinfo {pages} {050} (\bibinfo
  {year} {2016})},\ \Eprint {http://arxiv.org/abs/1507.06912} {arXiv:1507.06912
  [hep-ph]} \BibitemShut {NoStop}%
%%CITATION = ARXIV:1507.06912;%%
\bibitem [{\citenamefont {Iso}\ and\ \citenamefont
  {Okazawa}(2011)}]{Iso:2011gb}%
  \BibitemOpen
  \bibfield  {author} {\bibinfo {author} {\bibfnamefont {S.}~\bibnamefont
  {Iso}}\ and\ \bibinfo {author} {\bibfnamefont {S.}~\bibnamefont {Okazawa}},\
  }\href {\doibase 10.1016/j.nuclphysb.2011.05.021} {\bibfield  {journal}
  {\bibinfo  {journal} {Nucl. Phys.}\ }\textbf {\bibinfo {volume} {B851}},\
  \bibinfo {pages} {380} (\bibinfo {year} {2011})},\ \Eprint
  {http://arxiv.org/abs/1104.2461} {arXiv:1104.2461 [hep-th]} \BibitemShut
  {NoStop}%
%%CITATION = ARXIV:1104.2461;%%
\bibitem [{\citenamefont {Linde}, \citenamefont {Linde},\ and\ \citenamefont
  {Mezhlumian}(1994)}]{Linde:1993xx}%
  \BibitemOpen
  \bibfield  {author} {\bibinfo {author} {\bibfnamefont {A.~D.}\ \bibnamefont
  {Linde}}, \bibinfo {author} {\bibfnamefont {D.~A.}\ \bibnamefont {Linde}}, \
  and\ \bibinfo {author} {\bibfnamefont {A.}~\bibnamefont {Mezhlumian}},\
  }\href {\doibase 10.1103/PhysRevD.49.1783} {\bibfield  {journal} {\bibinfo
  {journal} {Phys. Rev.}\ }\textbf {\bibinfo {volume} {D49}},\ \bibinfo {pages}
  {1783} (\bibinfo {year} {1994})},\ \Eprint
  {http://arxiv.org/abs/gr-qc/9306035} {arXiv:gr-qc/9306035 [gr-qc]}
  \BibitemShut {NoStop}%
%%CITATION = GR-QC/9306035;%%
\bibitem [{\citenamefont {Linde}(1992)}]{Linde:1991sk}%
  \BibitemOpen
  \bibfield  {author} {\bibinfo {author} {\bibfnamefont {A.~D.}\ \bibnamefont
  {Linde}},\ }\href {\doibase 10.1016/0550-3213(92)90326-7} {\bibfield
  {journal} {\bibinfo  {journal} {Nucl. Phys.}\ }\textbf {\bibinfo {volume}
  {B372}},\ \bibinfo {pages} {421} (\bibinfo {year} {1992})},\ \Eprint
  {http://arxiv.org/abs/hep-th/9110037} {arXiv:hep-th/9110037 [hep-th]}
  \BibitemShut {NoStop}%
%%CITATION = HEP-TH/9110037;%%
\bibitem [{\citenamefont {Linde}(1979)}]{Linde:1978px}%
  \BibitemOpen
  \bibfield  {author} {\bibinfo {author} {\bibfnamefont {A.~D.}\ \bibnamefont
  {Linde}},\ }\href {\doibase 10.1088/0034-4885/42/3/001} {\bibfield  {journal}
  {\bibinfo  {journal} {Rept. Prog. Phys.}\ }\textbf {\bibinfo {volume} {42}},\
  \bibinfo {pages} {389} (\bibinfo {year} {1979})}\BibitemShut {NoStop}%
%%CITATION = RPPHA,42,389;%%
\bibitem [{\citenamefont {Dine}\ \emph {et~al.}(1992)\citenamefont {Dine},
  \citenamefont {Leigh}, \citenamefont {Huet}, \citenamefont {Linde},\ and\
  \citenamefont {Linde}}]{Dine:1992wr}%
  \BibitemOpen
  \bibfield  {author} {\bibinfo {author} {\bibfnamefont {M.}~\bibnamefont
  {Dine}}, \bibinfo {author} {\bibfnamefont {R.~G.}\ \bibnamefont {Leigh}},
  \bibinfo {author} {\bibfnamefont {P.~Y.}\ \bibnamefont {Huet}}, \bibinfo
  {author} {\bibfnamefont {A.~D.}\ \bibnamefont {Linde}}, \ and\ \bibinfo
  {author} {\bibfnamefont {D.~A.}\ \bibnamefont {Linde}},\ }\href {\doibase
  10.1103/PhysRevD.46.550} {\bibfield  {journal} {\bibinfo  {journal} {Phys.
  Rev.}\ }\textbf {\bibinfo {volume} {D46}},\ \bibinfo {pages} {550} (\bibinfo
  {year} {1992})},\ \Eprint {http://arxiv.org/abs/hep-ph/9203203}
  {arXiv:hep-ph/9203203 [hep-ph]} \BibitemShut {NoStop}%
%%CITATION = HEP-PH/9203203;%%
\bibitem [{\citenamefont {Candelas}(1980)}]{Candelas:1980zt}%
  \BibitemOpen
  \bibfield  {author} {\bibinfo {author} {\bibfnamefont {P.}~\bibnamefont
  {Candelas}},\ }\href {\doibase 10.1103/PhysRevD.21.2185} {\bibfield
  {journal} {\bibinfo  {journal} {Phys. Rev.}\ }\textbf {\bibinfo {volume}
  {D21}},\ \bibinfo {pages} {2185} (\bibinfo {year} {1980})}\BibitemShut
  {NoStop}%
%%CITATION = PHRVA,D21,2185;%%
\bibitem [{\citenamefont {Boulware}(1975{\natexlab{a}})}]{Boulware:1974dm}%
  \BibitemOpen
  \bibfield  {author} {\bibinfo {author} {\bibfnamefont {D.~G.}\ \bibnamefont
  {Boulware}},\ }\href {\doibase 10.1103/PhysRevD.11.1404} {\bibfield
  {journal} {\bibinfo  {journal} {Phys. Rev.}\ }\textbf {\bibinfo {volume}
  {D11}},\ \bibinfo {pages} {1404} (\bibinfo {year}
  {1975}{\natexlab{a}})}\BibitemShut {NoStop}%
%%CITATION = PHRVA,D11,1404;%%
\bibitem [{\citenamefont {Boulware}(1975{\natexlab{b}})}]{Boulware:1975pe}%
  \BibitemOpen
  \bibfield  {author} {\bibinfo {author} {\bibfnamefont {D.~G.}\ \bibnamefont
  {Boulware}},\ }\href {\doibase 10.1103/PhysRevD.12.350} {\bibfield  {journal}
  {\bibinfo  {journal} {Phys. Rev.}\ }\textbf {\bibinfo {volume} {D12}},\
  \bibinfo {pages} {350} (\bibinfo {year} {1975}{\natexlab{b}})}\BibitemShut
  {NoStop}%
%%CITATION = PHRVA,D12,350;%%
\bibitem [{\citenamefont {Unruh}(1976)}]{Unruh:1976db}%
  \BibitemOpen
  \bibfield  {author} {\bibinfo {author} {\bibfnamefont {W.~G.}\ \bibnamefont
  {Unruh}},\ }\href {\doibase 10.1103/PhysRevD.14.870} {\bibfield  {journal}
  {\bibinfo  {journal} {Phys. Rev.}\ }\textbf {\bibinfo {volume} {D14}},\
  \bibinfo {pages} {870} (\bibinfo {year} {1976})}\BibitemShut {NoStop}%
%%CITATION = PHRVA,D14,870;%%
\bibitem [{\citenamefont {Hartle}\ and\ \citenamefont
  {Hawking}(1976)}]{Hartle:1976tp}%
  \BibitemOpen
  \bibfield  {author} {\bibinfo {author} {\bibfnamefont {J.~B.}\ \bibnamefont
  {Hartle}}\ and\ \bibinfo {author} {\bibfnamefont {S.~W.}\ \bibnamefont
  {Hawking}},\ }\href {\doibase 10.1103/PhysRevD.13.2188} {\bibfield  {journal}
  {\bibinfo  {journal} {Phys. Rev.}\ }\textbf {\bibinfo {volume} {D13}},\
  \bibinfo {pages} {2188} (\bibinfo {year} {1976})}\BibitemShut {NoStop}%
%%CITATION = PHRVA,D13,2188;%%
\bibitem [{\citenamefont {DeWitt}(1975)}]{DeWitt:1975ys}%
  \BibitemOpen
  \bibfield  {author} {\bibinfo {author} {\bibfnamefont {B.~S.}\ \bibnamefont
  {DeWitt}},\ }\href {\doibase 10.1016/0370-1573(75)90051-4} {\bibfield
  {journal} {\bibinfo  {journal} {Phys. Rept.}\ }\textbf {\bibinfo {volume}
  {19}},\ \bibinfo {pages} {295} (\bibinfo {year} {1975})}\BibitemShut
  {NoStop}%
%%CITATION = PRPLC,19,295;%%
\bibitem [{\citenamefont {Christensen}\ and\ \citenamefont
  {Fulling}(1977)}]{Christensen:1977jc}%
  \BibitemOpen
  \bibfield  {author} {\bibinfo {author} {\bibfnamefont {S.~M.}\ \bibnamefont
  {Christensen}}\ and\ \bibinfo {author} {\bibfnamefont {S.~A.}\ \bibnamefont
  {Fulling}},\ }\href {\doibase 10.1103/PhysRevD.15.2088} {\bibfield  {journal}
  {\bibinfo  {journal} {Phys. Rev.}\ }\textbf {\bibinfo {volume} {D15}},\
  \bibinfo {pages} {2088} (\bibinfo {year} {1977})}\BibitemShut {NoStop}%
%%CITATION = PHRVA,D15,2088;%%
\bibitem [{\citenamefont {Christensen}(1976)}]{Christensen:1976vb}%
  \BibitemOpen
  \bibfield  {author} {\bibinfo {author} {\bibfnamefont {S.~M.}\ \bibnamefont
  {Christensen}},\ }\href {\doibase 10.1103/PhysRevD.14.2490} {\bibfield
  {journal} {\bibinfo  {journal} {Phys. Rev.}\ }\textbf {\bibinfo {volume}
  {D14}},\ \bibinfo {pages} {2490} (\bibinfo {year} {1976})}\BibitemShut
  {NoStop}%
%%CITATION = PHRVA,D14,2490;%%
\bibitem [{\citenamefont {Anderson}, \citenamefont {Hiscock},\ and\
  \citenamefont {Samuel}(1995)}]{Anderson:1994hg}%
  \BibitemOpen
  \bibfield  {author} {\bibinfo {author} {\bibfnamefont {P.~R.}\ \bibnamefont
  {Anderson}}, \bibinfo {author} {\bibfnamefont {W.~A.}\ \bibnamefont
  {Hiscock}}, \ and\ \bibinfo {author} {\bibfnamefont {D.~A.}\ \bibnamefont
  {Samuel}},\ }\href {\doibase 10.1103/PhysRevD.51.4337} {\bibfield  {journal}
  {\bibinfo  {journal} {Phys. Rev.}\ }\textbf {\bibinfo {volume} {D51}},\
  \bibinfo {pages} {4337} (\bibinfo {year} {1995})}\BibitemShut {NoStop}%
%%CITATION = PHRVA,D51,4337;%%
\bibitem [{\citenamefont {Page}(1982)}]{Page:1982fm}%
  \BibitemOpen
  \bibfield  {author} {\bibinfo {author} {\bibfnamefont {D.~N.}\ \bibnamefont
  {Page}},\ }\href {\doibase 10.1103/PhysRevD.25.1499} {\bibfield  {journal}
  {\bibinfo  {journal} {Phys. Rev.}\ }\textbf {\bibinfo {volume} {D25}},\
  \bibinfo {pages} {1499} (\bibinfo {year} {1982})}\BibitemShut {NoStop}%
%%CITATION = PHRVA,D25,1499;%%
\bibitem [{\citenamefont {Brown}, \citenamefont {Ottewill},\ and\ \citenamefont
  {Page}(1986)}]{Brown:1986jy}%
  \BibitemOpen
  \bibfield  {author} {\bibinfo {author} {\bibfnamefont {M.~R.}\ \bibnamefont
  {Brown}}, \bibinfo {author} {\bibfnamefont {A.~C.}\ \bibnamefont {Ottewill}},
  \ and\ \bibinfo {author} {\bibfnamefont {D.~N.}\ \bibnamefont {Page}},\
  }\href {\doibase 10.1103/PhysRevD.33.2840} {\bibfield  {journal} {\bibinfo
  {journal} {Phys. Rev.}\ }\textbf {\bibinfo {volume} {D33}},\ \bibinfo {pages}
  {2840} (\bibinfo {year} {1986})}\BibitemShut {NoStop}%
%%CITATION = PHRVA,D33,2840;%%
\bibitem [{\citenamefont {Frolov}\ and\ \citenamefont
  {Zelnikov}(1987)}]{Frolov:1987gw}%
  \BibitemOpen
  \bibfield  {author} {\bibinfo {author} {\bibfnamefont {V.~P.}\ \bibnamefont
  {Frolov}}\ and\ \bibinfo {author} {\bibfnamefont {A.~I.}\ \bibnamefont
  {Zelnikov}},\ }\href {\doibase 10.1103/PhysRevD.35.3031} {\bibfield
  {journal} {\bibinfo  {journal} {Phys. Rev.}\ }\textbf {\bibinfo {volume}
  {D35}},\ \bibinfo {pages} {3031} (\bibinfo {year} {1987})}\BibitemShut
  {NoStop}%
%%CITATION = PHRVA,D35,3031;%%
\bibitem [{\citenamefont {Vaz}(1989)}]{Vaz:1988gh}%
  \BibitemOpen
  \bibfield  {author} {\bibinfo {author} {\bibfnamefont {C.}~\bibnamefont
  {Vaz}},\ }\href {\doibase 10.1103/PhysRevD.39.1776} {\bibfield  {journal}
  {\bibinfo  {journal} {Phys. Rev.}\ }\textbf {\bibinfo {volume} {D39}},\
  \bibinfo {pages} {1776} (\bibinfo {year} {1989})}\BibitemShut {NoStop}%
%%CITATION = PHRVA,D39,1776;%%
\bibitem [{\citenamefont {Barrios}\ and\ \citenamefont
  {Vaz}(1989)}]{Barrios:1990vg}%
  \BibitemOpen
  \bibfield  {author} {\bibinfo {author} {\bibfnamefont {F.~A.}\ \bibnamefont
  {Barrios}}\ and\ \bibinfo {author} {\bibfnamefont {C.}~\bibnamefont {Vaz}},\
  }\href {\doibase 10.1103/PhysRevD.40.1340} {\bibfield  {journal} {\bibinfo
  {journal} {Phys. Rev.}\ }\textbf {\bibinfo {volume} {D40}},\ \bibinfo {pages}
  {1340} (\bibinfo {year} {1989})}\BibitemShut {NoStop}%
%%CITATION = PHRVA,D40,1340;%%
\bibitem [{\citenamefont {Howard}\ and\ \citenamefont
  {Candelas}(1984)}]{Howard:1984qp}%
  \BibitemOpen
  \bibfield  {author} {\bibinfo {author} {\bibfnamefont {K.~W.}\ \bibnamefont
  {Howard}}\ and\ \bibinfo {author} {\bibfnamefont {P.}~\bibnamefont
  {Candelas}},\ }\href {\doibase 10.1103/PhysRevLett.53.403} {\bibfield
  {journal} {\bibinfo  {journal} {Phys. Rev. Lett.}\ }\textbf {\bibinfo
  {volume} {53}},\ \bibinfo {pages} {403} (\bibinfo {year} {1984})}\BibitemShut
  {NoStop}%
%%CITATION = PRLTA,53,403;%%
\bibitem [{\citenamefont {Howard}(1984)}]{Howard:1985yg}%
  \BibitemOpen
  \bibfield  {author} {\bibinfo {author} {\bibfnamefont {K.~W.}\ \bibnamefont
  {Howard}},\ }\href {\doibase 10.1103/PhysRevD.30.2532} {\bibfield  {journal}
  {\bibinfo  {journal} {Phys. Rev.}\ }\textbf {\bibinfo {volume} {D30}},\
  \bibinfo {pages} {2532} (\bibinfo {year} {1984})}\BibitemShut {NoStop}%
%%CITATION = PHRVA,D30,2532;%%
\bibitem [{\citenamefont {Nugaev}(1991)}]{Nugaev:1991aa}%
  \BibitemOpen
  \bibfield  {author} {\bibinfo {author} {\bibfnamefont {R.~M.}\ \bibnamefont
  {Nugaev}},\ }\href {\doibase 10.1103/PhysRevD.43.1195} {\bibfield  {journal}
  {\bibinfo  {journal} {Phys. Rev.}\ }\textbf {\bibinfo {volume} {D43}},\
  \bibinfo {pages} {1195} (\bibinfo {year} {1991})}\BibitemShut {NoStop}%
%%CITATION = PHRVA,D43,1195;%%
\bibitem [{\citenamefont {Matyjasek}(1996{\natexlab{a}})}]{Matyjasek:1996dm}%
  \BibitemOpen
  \bibfield  {author} {\bibinfo {author} {\bibfnamefont {J.}~\bibnamefont
  {Matyjasek}},\ }\href {\doibase 10.1103/PhysRevD.53.794} {\bibfield
  {journal} {\bibinfo  {journal} {Phys. Rev.}\ }\textbf {\bibinfo {volume}
  {D53}},\ \bibinfo {pages} {794} (\bibinfo {year}
  {1996}{\natexlab{a}})}\BibitemShut {NoStop}%
%%CITATION = PHRVA,D53,794;%%
\bibitem [{\citenamefont {Matyjasek}(1996{\natexlab{b}})}]{Matyjasek:1996if}%
  \BibitemOpen
  \bibfield  {author} {\bibinfo {author} {\bibfnamefont {J.}~\bibnamefont
  {Matyjasek}},\ }\href@noop {} {\bibfield  {journal} {\bibinfo  {journal}
  {Acta Phys. Polon.}\ }\textbf {\bibinfo {volume} {B27}},\ \bibinfo {pages}
  {2005} (\bibinfo {year} {1996}{\natexlab{b}})}\BibitemShut {NoStop}%
%%CITATION = APPOA,B27,2005;%%
\bibitem [{\citenamefont {Matyjasek}(1997{\natexlab{a}})}]{Matyjasek:1996ig}%
  \BibitemOpen
  \bibfield  {author} {\bibinfo {author} {\bibfnamefont {J.}~\bibnamefont
  {Matyjasek}},\ }\href {\doibase 10.1088/0264-9381/14/1/003} {\bibfield
  {journal} {\bibinfo  {journal} {Class. Quant. Grav.}\ }\textbf {\bibinfo
  {volume} {14}},\ \bibinfo {pages} {L15} (\bibinfo {year}
  {1997}{\natexlab{a}})}\BibitemShut {NoStop}%
%%CITATION = CQGRD,14,L15;%%
\bibitem [{\citenamefont {Matyjasek}(1997{\natexlab{b}})}]{Matyjasek:1996ih}%
  \BibitemOpen
  \bibfield  {author} {\bibinfo {author} {\bibfnamefont {J.}~\bibnamefont
  {Matyjasek}},\ }\href {\doibase 10.1103/PhysRevD.55.809} {\bibfield
  {journal} {\bibinfo  {journal} {Phys. Rev.}\ }\textbf {\bibinfo {volume}
  {D55}},\ \bibinfo {pages} {809} (\bibinfo {year}
  {1997}{\natexlab{b}})}\BibitemShut {NoStop}%
%%CITATION = PHRVA,D55,809;%%
\bibitem [{\citenamefont {Visser}(1997)}]{Visser:1997sd}%
  \BibitemOpen
  \bibfield  {author} {\bibinfo {author} {\bibfnamefont {M.}~\bibnamefont
  {Visser}},\ }\href {\doibase 10.1103/PhysRevD.56.936} {\bibfield  {journal}
  {\bibinfo  {journal} {Phys. Rev.}\ }\textbf {\bibinfo {volume} {D56}},\
  \bibinfo {pages} {936} (\bibinfo {year} {1997})},\ \Eprint
  {http://arxiv.org/abs/gr-qc/9703001} {arXiv:gr-qc/9703001 [gr-qc]}
  \BibitemShut {NoStop}%
%%CITATION = GR-QC/9703001;%%
\bibitem [{\citenamefont {Matyjasek}(1998)}]{Matyjasek:1998zs}%
  \BibitemOpen
  \bibfield  {author} {\bibinfo {author} {\bibfnamefont {J.}~\bibnamefont
  {Matyjasek}},\ }\href {\doibase 10.1103/PhysRevD.57.7615} {\bibfield
  {journal} {\bibinfo  {journal} {Phys. Rev.}\ }\textbf {\bibinfo {volume}
  {D57}},\ \bibinfo {pages} {7615} (\bibinfo {year} {1998})}\BibitemShut
  {NoStop}%
%%CITATION = PHRVA,D57,7615;%%
\bibitem [{\citenamefont {Matyjasek}(1999)}]{Matyjasek:1998mq}%
  \BibitemOpen
  \bibfield  {author} {\bibinfo {author} {\bibfnamefont {J.}~\bibnamefont
  {Matyjasek}},\ }\href {\doibase 10.1103/PhysRevD.59.044002} {\bibfield
  {journal} {\bibinfo  {journal} {Phys. Rev.}\ }\textbf {\bibinfo {volume}
  {D59}},\ \bibinfo {pages} {044002} (\bibinfo {year} {1999})},\ \Eprint
  {http://arxiv.org/abs/gr-qc/9808019} {arXiv:gr-qc/9808019 [gr-qc]}
  \BibitemShut {NoStop}%
%%CITATION = GR-QC/9808019;%%
\bibitem [{\citenamefont {Ford}\ \emph {et~al.}(1993)\citenamefont {Ford},
  \citenamefont {Jones}, \citenamefont {Stephenson},\ and\ \citenamefont
  {Einhorn}}]{Ford:1992mv}%
  \BibitemOpen
  \bibfield  {author} {\bibinfo {author} {\bibfnamefont {C.}~\bibnamefont
  {Ford}}, \bibinfo {author} {\bibfnamefont {D.~R.~T.}\ \bibnamefont {Jones}},
  \bibinfo {author} {\bibfnamefont {P.~W.}\ \bibnamefont {Stephenson}}, \ and\
  \bibinfo {author} {\bibfnamefont {M.~B.}\ \bibnamefont {Einhorn}},\ }\href
  {\doibase 10.1016/0550-3213(93)90206-5} {\bibfield  {journal} {\bibinfo
  {journal} {Nucl. Phys.}\ }\textbf {\bibinfo {volume} {B395}},\ \bibinfo
  {pages} {17} (\bibinfo {year} {1993})},\ \Eprint
  {http://arxiv.org/abs/hep-lat/9210033} {arXiv:hep-lat/9210033 [hep-lat]}
  \BibitemShut {NoStop}%
%%CITATION = HEP-LAT/9210033;%%
\bibitem [{\citenamefont {Casas}, \citenamefont {Espinosa},\ and\ \citenamefont
  {Quiros}(1995)}]{Casas:1994qy}%
  \BibitemOpen
  \bibfield  {author} {\bibinfo {author} {\bibfnamefont {J.~A.}\ \bibnamefont
  {Casas}}, \bibinfo {author} {\bibfnamefont {J.~R.}\ \bibnamefont {Espinosa}},
  \ and\ \bibinfo {author} {\bibfnamefont {M.}~\bibnamefont {Quiros}},\ }\href
  {\doibase 10.1016/0370-2693(94)01404-Z} {\bibfield  {journal} {\bibinfo
  {journal} {Phys. Lett.}\ }\textbf {\bibinfo {volume} {B342}},\ \bibinfo
  {pages} {171} (\bibinfo {year} {1995})},\ \Eprint
  {http://arxiv.org/abs/hep-ph/9409458} {arXiv:hep-ph/9409458 [hep-ph]}
  \BibitemShut {NoStop}%
%%CITATION = HEP-PH/9409458;%%
\bibitem [{\citenamefont {Espinosa}\ and\ \citenamefont
  {Quiros}(1995)}]{Espinosa:1995se}%
  \BibitemOpen
  \bibfield  {author} {\bibinfo {author} {\bibfnamefont {J.~R.}\ \bibnamefont
  {Espinosa}}\ and\ \bibinfo {author} {\bibfnamefont {M.}~\bibnamefont
  {Quiros}},\ }\href {\doibase 10.1016/0370-2693(95)00572-3} {\bibfield
  {journal} {\bibinfo  {journal} {Phys. Lett.}\ }\textbf {\bibinfo {volume}
  {B353}},\ \bibinfo {pages} {257} (\bibinfo {year} {1995})},\ \Eprint
  {http://arxiv.org/abs/hep-ph/9504241} {arXiv:hep-ph/9504241 [hep-ph]}
  \BibitemShut {NoStop}%
%%CITATION = HEP-PH/9504241;%%
\bibitem [{\citenamefont {Gregory}, \citenamefont {Moss},\ and\ \citenamefont
  {Withers}(2014)}]{Gregory:2013hja}%
  \BibitemOpen
  \bibfield  {author} {\bibinfo {author} {\bibfnamefont {R.}~\bibnamefont
  {Gregory}}, \bibinfo {author} {\bibfnamefont {I.~G.}\ \bibnamefont {Moss}}, \
  and\ \bibinfo {author} {\bibfnamefont {B.}~\bibnamefont {Withers}},\ }\href
  {\doibase 10.1007/JHEP03(2014)081} {\bibfield  {journal} {\bibinfo  {journal}
  {JHEP}\ }\textbf {\bibinfo {volume} {03}},\ \bibinfo {pages} {081} (\bibinfo
  {year} {2014})},\ \Eprint {http://arxiv.org/abs/1401.0017} {arXiv:1401.0017
  [hep-th]} \BibitemShut {NoStop}%
%%CITATION = ARXIV:1401.0017;%%
\bibitem [{\citenamefont {Carr}\ \emph {et~al.}(2010)\citenamefont {Carr},
  \citenamefont {Kohri}, \citenamefont {Sendouda},\ and\ \citenamefont
  {Yokoyama}}]{Carr:2009jm}%
  \BibitemOpen
  \bibfield  {author} {\bibinfo {author} {\bibfnamefont {B.~J.}\ \bibnamefont
  {Carr}}, \bibinfo {author} {\bibfnamefont {K.}~\bibnamefont {Kohri}},
  \bibinfo {author} {\bibfnamefont {Y.}~\bibnamefont {Sendouda}}, \ and\
  \bibinfo {author} {\bibfnamefont {J.}~\bibnamefont {Yokoyama}},\ }\href
  {\doibase 10.1103/PhysRevD.81.104019} {\bibfield  {journal} {\bibinfo
  {journal} {Phys. Rev.}\ }\textbf {\bibinfo {volume} {D81}},\ \bibinfo {pages}
  {104019} (\bibinfo {year} {2010})},\ \Eprint {http://arxiv.org/abs/0912.5297}
  {arXiv:0912.5297 [astro-ph.CO]} \BibitemShut {NoStop}%
%%CITATION = ARXIV:0912.5297;%%
\end{thebibliography}%
%%%%%%%%%%%%%%%%%%%%%%%%%%%%%%%%%%%%%%%%%%%%%%%%%%%%%%%%%%%%%%%%%%%%
\end{document}